\pdfoutput=1
\documentclass{sig-alternate-05-2015}

\usepackage{amsmath}    
\usepackage{amssymb}
\usepackage{wasysym}
\usepackage{mathrsfs}
\usepackage{graphicx}
\usepackage{subfigure}
\usepackage{enumerate}

\newcommand{\argmax}[1]{\underset{#1}{\mbox{arg max }}}

\newcommand{\script}[1]{{{\cal{#1} }}}

\newtheorem{lemma}{Lemma}

\newtheorem{assumption}{Assumption}
\newtheorem{prop}{Proposition}
\newtheorem{theorem}{Theorem}
\newtheorem{example}{Example}

\allowdisplaybreaks
\makeatletter
\def\@copyrightspace{\relax}
\makeatother
\begin{document}

\title{Prices and Subsidies in the Sharing Economy}

\numberofauthors{3}
\author{
\alignauthor
Zhixuan Fang\\
    \affaddr{Tsinghua Universiy }\\
    \affaddr{Beijing, China}
    \email{fzx13}\\
    \email{@mails.tsinghua.edu.cn}
\alignauthor
Longbo Huang\\
    \affaddr{Tsinghua Universiy}\\
    \affaddr{Beijing, China}\\
    \email{longbohuang}\\
    \email{@tsinghua.edu.cn}
\alignauthor
Adam Wierman\\
    \affaddr{California Institute of Technology}\\
    \affaddr{Pasadena, CA, USA}\\
    \email{adamw@caltech.edu}
}

\maketitle

\begin{abstract}
The growth of the sharing economy is driven by the emergence of sharing platforms, e.g., Uber and Lyft, that match owners looking to share their resources with customers looking to rent them.  The design of such platforms is a complex mixture of economics and engineering, and how to ``optimally'' design such platforms is still an open problem.  In this paper, we focus on the design of prices and subsidies in sharing platforms. Our results provide insights into the tradeoff between revenue maximizing prices and social welfare maximizing prices.  Specifically, we introduce a novel model of sharing platforms and characterize the profit and social welfare maximizing prices in this model.  Further, we bound the efficiency loss under profit maximizing prices, showing that there is a strong alignment between profit and efficiency in practical settings.  Our results highlight that the revenue of platforms may be limited in practice due to supply shortages; thus platforms have a strong incentive to encourage sharing via subsidies.  We provide an analytic characterization of when such subsidies are valuable and show how to optimize the size of the subsidy provided.  Finally, we validate the insights from our analysis using data from Didi Chuxing, the largest ridesharing platform in China.
\end{abstract}

\section{Introduction}

The growth of the sharing economy is driven by the emergence of sharing platforms that facilitate exchange, e.g., Uber, Lyft, and Didi Chuxing.
While initially limited to a few industries, e.g., ridesharing, sharing platforms have now emerged in diverse areas including household tasks, e.g., TaskRabbit, rental housing, e.g., Airbnb, HomeAway; food delivery, e.g., UberEATS, and more.

The sharing economies governed by these platforms are inherently two-sided markets: owners on one side look to rent access to some product and renters/customers on the other side look to use that product for a limited time period. The role of sharing platforms in this two-sided market is to facilitate matches between those looking to share and those looking to rent, a process that would be very difficult without a central exchange.

Since sharing platforms do not own their own products, their goal is mainly to act as an intermediary that ensures the availability of supply (shared resources), while maintaining high satisfaction in transactions that happen through the platform.  The determination of prices and matches by a sharing platform is the key mechanism for accomplishing this goal -- and it is a difficult, complex task that has significant impact on profit of the platform and satisfaction of the users, both owners and renters.

Sharing platforms today are diverse, both in how matches are assigned and in how prices are determined. Even within a single industry, e.g., ridesharing, platforms use contrasting approaches for pricing and matching.
For example, Uber slashed the fares  in US in January 2016 and began to  subsidize drivers heavily, while Lyft remained a relatively higher price option \cite{uber_subsidy}. These varied approaches to pricing highlight the limited understanding currently available about the impact of different approaches on profit and user satisfaction -- companies do not agree on an ``optimal'' approach.

This lack of understanding exists despite a large literature in economics studying two-sided markets, e.g., \cite{rysman2009economics, weyl2010price, rochet2006two}. This gap stems from the fact that, while sharing platforms are examples of two-sided markets, the traditional research on two-sided markets often does not apply to sharing platforms due to assumptions made about the form of utilities and interactions across the markets. Sharing platforms are faced with asymmetries between renters and owners, as well as non-traditional utility functions as a result of trading off the benefits of sharing with the benefits of personal usage.

This paper seeks to provide new insight into two related, important design choices in sharing platforms: \emph{the design of prices and subsidies}.

How a sharing platform sets prices has a crucial impact on both the availability of shared resources and the demand from customers for the shared resources.  A platform must provide incentives for sharing that ensure enough supply, while also keeping prices low enough that customers are willing to pay for the service.

This tradeoff is often difficult for sharing platforms to satisfy, especially in their early stages, and sharing platforms typically find themselves fighting to avoid supply shortages by giving subsidies for sharing.  In fact, according to \cite{didi_subsidy}, Didi spends up to $\$4$ billion  on subsidies per year.

Not surprisingly, sharing platforms carefully optimize prices and subsidies, often keeping the details of their approaches secret.  However, it is safe to assume that sharing platforms typically approach the design question with an eye on optimizing revenue obtained by the platform (via transaction fees).  In contrast, a driving motivation for the sharing economy is to encourage efficient resource pooling and thus achieving social welfare improvements through better utilization of resources.  Thus, a natural question is ``\emph{Do revenue maximizing sharing platforms (nearly) optimize the social welfare achievable through resource pooling?}"  Said differently, ``Do sharing platforms achieve market efficiency or is there inefficiency created by revenue-seeking platforms?''

\paragraph*{Contributions of this paper}  \emph{This paper adapts classical model of two-sided markets to the case of sharing platforms and uses this new model to guide the design of prices and subsidies in sharing platforms.}  The paper makes four main contributions to the literatures on two-sided markets and sharing platforms.

\emph{First}, this paper introduces a novel model of a two-sided market within the sharing economy.  Our model consists of a sharing platform, a set of product owners who are interested in sharing the product, and a set of product users (renters).
In classic two sided markets, users on both sides benefit from the increasing size of the other side, e.g., their benefits are linear to the size of the others. Maximizing the size of the market is the key problem for the platform.
Our model is different from the classic two-sided model in that we consider asymmetric market, i.e., product owners can benefit from either  renting their resource or using it themselves.

\emph{Second}, within our novel two-sided market model of a sharing platform, we prove existence and uniqueness of a Nash equilibrium and derive structural properties of market behavior as a function of the pricing strategy used by the sharing platform.  These results allow the characterization and comparison of pricing strategies that (i) maximize revenue and (ii) maximize social welfare. Our results highlight that revenue maximizing prices are always at least as large as social welfare maximizing prices (Theorem \ref{thm:pr_psw}) and, further, that the welfare loss from  revenue maximization is small (Theorem \ref{thm:gap}).  Thus, revenue-maximizing sharing platforms achieve nearly all of the gains possible from resource pooling through the sharing economy.  Interestingly, our results also highlight that revenue maximizing prices lead to more sharing (higher supply) compared to welfare maximizing pricing. Finally, perhaps counter-intuitively, our results show that revenue maximizing prices and social welfare maximizing prices align in situations where the market is ``congested'', i.e., where the sharing supply is low.  This is the operating regime of many sharing platforms, and so our results suggest that sharing platforms are likely operating in a regime where business and societal goals are aligned.

\emph{Third}, we provide results characterizing the impact of subsidies for sharing, and derive the optimal subsidies for maximizing revenue.  The fact that many sharing platforms are operating in ``congested'' regimes means that it is crucial for them to find ways to encourage sharing, and the most common approach is to subsidize sharing.  Theorem \ref{thm:subsidy} characterizes the market equilibrium as a function of the subsidy provided by the platform.  These results thus allow the platform to choose a subsidy that optimizes revenue (or some other objectives), including the costs of the subsidies themselves.  Importantly, the results highlight that small subsidies can have a dramatic impact on the available supply.

Finally, the \emph{fourth} contribution of the paper is an empirical exploration of a sharing platform in order to ground the theoretical work in the paper (Section \ref{sec:simulation}).  In particular, we use data from Didi Chuxing, the largest ridesharing platform in China, to fit and validate our model, and also to explore the insights from our theorems in practical settings.  Our results highlight that, in practical settings, revenue maximizing pricing shows close alignment with welfare maximizing pricing.  Further, we  explore the impact of supply-side regulation (e.g., gas taxes) as a method for curbing over-supply in sharing platforms.  Our results show that such approaches can have an impact, and that the reduction is felt entirely by the product owners (not the sharing platform).

\paragraph*{Related literature}

Our paper is related to two distinct literatures: (i) empirical work studying the sharing economy, and (ii) analytical work studying two-sided markets.  

\textbf{Empirical studies of the sharing economy.} There is a growing literature studying the operation of the sharing economy.  Much of this work is empirical, focused on quantifying the benefits and drawbacks of the sharing economy \cite{dillahunt2015promise, Fang2016264, malhotra2014dark}, the operation of existing sharing platforms \cite{cohen2016using,zervas2015first}, and the social consequences of current designs \cite{quattrone2016benefits,zervas2014rise}.

An important insight from these literature relevant to the model in the current paper is that the expected economic benefits of owners due to sharing significantly influence the level of participation in sharing platforms \cite{hamari2015sharing, lamberton2012ours}.  Thus, prices and subsidies do impact the degree to which owners participate in the sharing economy.  Similarly, studies have shown that prices have a dramatic impact on demand in sharing platforms, e.g., \cite{chen2015peeking} conducts experiments in San Francisco and Manhattan to show that Uber's surge pricing dramatically decreases demand.

\textbf{Analytical studies of two-sided markets.}  There is a large literature on two-sided markets in the economics community, e.g., see \cite{rysman2009economics, weyl2010price, rochet2006two} and the references therein for an overview.  These papers typically focus on situations where users on one side of the market benefit from participation on the other side of the market.  Hence, the goal of the platform is to increase the population on both sides.  However, the models considered in this literature do not apply to sharing platforms, as we discuss in detail in Section \ref{sec:model}.

That said, there is a small, but growing, set of papers that attempts to adapt models of two-sided markets in order to study sharing platforms.  These papers tend to focus on a specific feature of a specific sharing platform, e.g., dynamic pricing for ridesharing.  Most related to the current paper are
\cite{benjaafar2015peer} and \cite{jiang2015collaborative}, which  study  platform strategies and  social welfare assuming users have fixed usage value. Another related paper is \cite{einav2015peer}, which compares the utility of dedicated and flexible sellers, showing that cost is the key factor that influences participation of sellers. Finally, \cite{cachon2015role} analyzes competition among providers, and shows that the commonly adopted commission contract, i.e., platform extracts commission fee at a fixed ratio from transaction value, is nearly optimal.

Our work differs significantly from each of the above mentioned papers.  In particular, we consider heterogenous users with general concave utility functions, while prior works often consider only linear utilities or adopt a specific form of utility functions among homogeneous users. Additionally, in our model, each user's private information is not revealed to others or the platform, while many results in prior work is based on known knowledge of user utilities. Further, unlike the agents in traditional models of two-sided markets, product providers in sharing platform can benefit from either self-usage of the products or sharing. This asymmetry significantly impacts the analysis and results. Finally, our work is unique among the analytic studies of sharing platforms in that we use data from Didi Chuxing to fit our model and validate the insights.

\section{Modeling a sharing platform}\label{sec:model}

This paper seeks to provide insight into the design of prices and subsidies in a sharing platform.  To that end, we begin by presenting a novel analytic model for the interaction of agents within a sharing platform.  The model is a variation of traditional models for two-sided markets, e.g., \cite{rochet2006two,weyl2010price}, that includes important adjustments to capture the asymmetries created by interactions between owners and renters in a sharing platform.  To ground our modeling, we discuss the model in terms of ride-sharing and use data from Didi Chuxing \cite{didi} to guide our modeling choices.

\subsection{Model preliminaries}

We consider a marketplace consisting of a \emph{sharing platform} and two groups of users: (i) \emph{owners}, denoted by the set $\mathcal{O}$, and (ii) \emph{renters}, denoted by $\mathcal{R}$.  For example, in a ridesharing platform, owners are those that sometimes use their car themselves and other times rent their car through the platform, e.g., Uber or Didi Chuxing, and renters are those that use the platform to get rides.

We use $N_O = |\mathcal{O}|$ and $N_R=|\mathcal{R}|$ to denote  the number of owners and renters, respectively, and assume that $N_O \geq 2$.\footnote{The case when $N_O=1$ is the monopoly case. It is not the main focus of this paper and can be analyzed in a straightforward manner (though the argument is different from the one used for the case discussed in this paper).} For each user in $\mathcal{O}\cup \mathcal{R}$, we normalize the maximum usage to be $1$.  This should be thought of as the product usage frequency. For product owner $i$, let $x_i\in[0,1]$ denote the self-usage level, e.g., the fraction of time the owner uses the product personally.  Further, let $s_i\in[0, 1]$ denote the level at which the owner \emph{shares} the product on the platform, e.g., drives his or her car in a ridesharing service. Note that we always have $x_i+s_i\leq 1$.
For a renter $k\in\mathcal{R}$, we use $y_k\in[0, 1]$ to denote the usage level of the product, e.g., the fraction of time when using a ridesharing service.

Sharing takes place over the platform. In our model, the sharing platform first sets a (homogenous) market price $p$.\footnote{We consider homogeneous prices for simplicity of exposition.  One could also consider heterogeneous prices by introducing different types of owners and renters and parameterizing their utility functions appropriately. The structure of the model and results remain in such an extension.} Think of this price as a price per time for renting a product through the platform. For example, prices in the Didi data set we describe in Section \ref{sec:simulation} are approximately affine in length.\footnote{Note that ridesharing platforms typically use \emph{dynamic} pricing.  Dynamic pricing is not the focus of our paper.  Our focus is on determining optimal prices and subsidies for a given time. Thus, our model considers only a static point in time and is not appropriate for a study of dynamic pricing.  Interested readers should refer to \cite{Banerjee:2015} for a complementary, recent work specifically focused on dynamic pricing in ridesharing platforms. Interestingly, \cite{Banerjee:2015} shows that dynamic pricing does not yield higher efficiency than static pricing, though it does provide improved robustness.}

Renters pay the market price for renting a product from owners.  Depending on the self-usage preferences and the market price, owners choose their own $x_i$ and $s_i$  values ($y_k$ for product renters), which together result in different demand and supply conditions in the system.  We describe the models that govern these choices in the subsections below.

The aggregate choices are summarized by the following notation: $S(p)$ is the total sharing supply and $D(p)$ the total demand (renting demand) in the market under price $p$, i.e.,
\begin{eqnarray}
S(p)=\sum_{i\in \mathcal{O}}s_{i}(p),\quad
D(p)=\sum_{k\in \mathcal{R}}y_{k}(p).
\end{eqnarray}

With the notation and terminology set, we now focus on the detailed models of the strategic interactions between owners, renters, and the platform.

\subsection{Modeling renters}
For each user $k\in\mathcal{R}$, we denote the usage benefit obtained from using the product, e.g., convenience or personal satisfaction,  by $g_k(y_k)$.
We assume that $g_k$ is only known to the renter and is not public knowledge.

The utility a renter $k$ obtains from product usage $y_k$, $U_{k}(y_{k})$, is given by the benefit $g_k$ minus the cost to rent the product, $p$, times the usage $y_k$. Hence,
\begin{eqnarray}\label{eq:utilty_renter}
U_{k}(y_{k})=g_{k}(y_{k})-py_{k}.
\end{eqnarray}
Naturally, each renter $k$ chooses demand $y_{k}$ to maximize utility, i.e.,
\vspace{-.06in}
\begin{eqnarray}
\max_{y_k} \quad U_k(y_{k})\quad \text{s.t.} \quad 0\leq y_{k}\leq 1. \label{eq:optmize-renter}
\end{eqnarray}
For analytic reasons, we assume that $g_k(y)$ is continuous and strictly concave  with $g_{k}(0)=0$. Moreover, we assume that $\partial_+g_k(0)\leq B$ for some $B>0$, where $\partial_{+}g_k(0)$ is the right derivative of $g_k(y)$ at $y=0$.  Note that  concavity, continuity, and bounded derivatives are standard assumptions in the utility analysis literature, e.g., \cite{rosen1965, rabin2000risk}.

\begin{figure}
\centering
\vspace{-0.1in}
\includegraphics[scale=0.20]{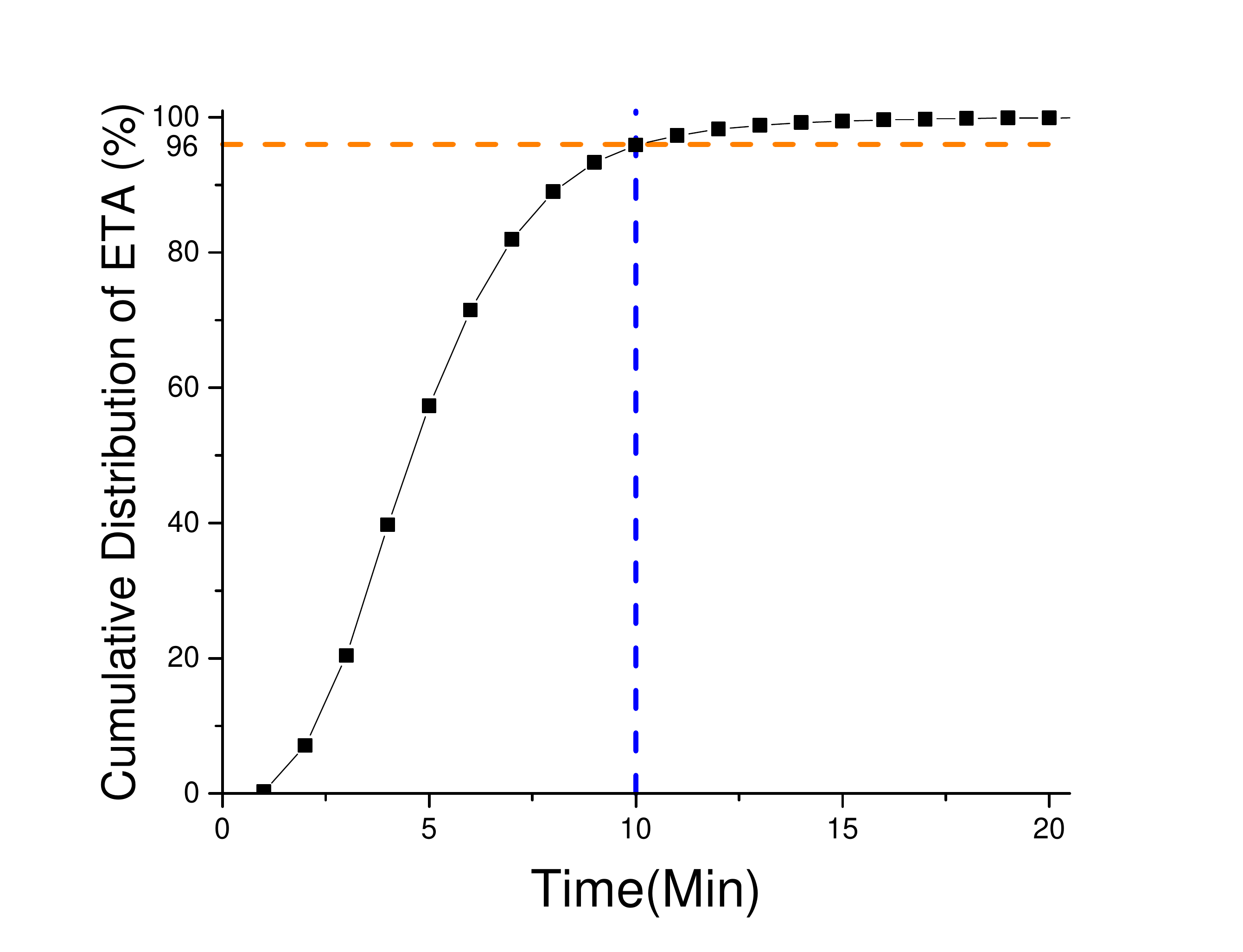}
\vspace{-0.2in}
\caption{Waiting time is typically short: average ETA of UberX in Washington D.C. is $5.5$min and $96\%$ requests has ETA less than $10$min, which is much better than taxi \cite{uber-waittime}.
}
\label{fig:eta}
\end{figure}

The key assumption in the model above is that we are modeling the choice of renters as being primarily a function of price.  This choice is motivated by data from ridesharing services such as Uber and Didi.  In particular, one may wonder if the impact of price is modulated by other factors, e.g., estimated time of arrival (ETA).  However, data indicates that these other factors play a much smaller role. This is because, in most cases, the ETA is small, and hence differences in ETA are less salient than differences in price.
For instance, Figure \ref{fig:eta}  shows the real ETA  statistics of UberX in the Washington D.C. area for $6.97$ million requests (from dataset \cite{uber-waittime}). Notice that the average waiting time is $5.5$ minutes and $96\%$ of requests have an ETA less than $10$ minutes. Within that scale, wait times do not have a major impact on user behavior \cite{wait_time}.

In contrast, price fluctuations have a significant impact on user behavior.  To highlight this, consider Figure \ref{fig:data}, which shows data we have obtained from 395,938 transactions in Didi.  (We introduce and explore this data set in detail in Section \ref{sec:simulation}.)  This data set highlights that price fluctuations have a significant impact on user behavior.  Further, it can be used to reverse engineer a realistic model of what renter utility functions may look like in practice.  In particular, the following concave form fits the data well (see Section \ref{sec:simulation} for more details):
\vspace{-.06in}
\begin{equation*}
g_k(y_k)=\frac{1}{\beta}(y_k+y_k\log(\frac{\alpha}{N_R})-y_k\log y_k).
\end{equation*}

Beyond ridesharing, price can also be seen to play the primary role in renter decision making in other sharing economy platforms as well.  For example, in Airbnb, since destination cities and dates are usually subject to travelers' schedule, price becomes a main concern for renters.

\subsection{Modeling owners}
The key distinction between owners and renters is that owners have two ways to derive benefits from the product -- using it themselves or renting it  through the platform.

When owners use the product themselves, they experience benefit from the usage, like renters.  But, unlike renters, they do not have to pay the platform for their usage, though their usage does incur wear-and-tear and thus leads to maintenance costs.  We denote the benefit from  self-usage by $f_{i}(x_{i})$ and the maintenance costs incurred by $cx_i$, where $c\geq 0$ is a constant representing the usage cost per unit, e.g., gasoline, house keeping, or government tax.  As in the case of renters, we assume that, for all owners, $f_i(\cdot)$ is continuous and strictly concave  with  $f_{i}(0)=0$ and $\partial_+f_i(0)\leq B$.

When the owners share their product through the platform, they receive income; however they also incur costs.  Crucially, the income they receive from sharing depends on how many renters are present and how many other owners are sharing, and thus competing for renters.

We model the competition between owners via the following simple equation for the income received while sharing:
\begin{equation}p\min\{\frac{D(p)}{S(p)},1\}s_{i}.\end{equation}
Here $\min\{\frac{D(p)}{S(p)},1\}$ denotes the probability that an owner is matched to a renter.  This models a ``fair'' platform that is equally likely to assign a renter to any owner in the system. Note that fairness among owners is crucial for encouraging participation in platforms.  In a ``fair'' platform $\min\{\frac{D(p)}{S(p)},1\}$ is the fraction of time owner $i$ makes money from sharing, e.g., the fraction of time a driver in Uber has a rider.  Thus, $p\min\{\frac{D(p)}{S(p},1\}$ can be viewed as the revenue stream that an owner sees while sharing.\footnote{Similar utility functions have been adopted  for studying resources allocation problems with symmetric users in other contexts, e.g., \cite{Even-dar:2009} and \cite{johari2009efficiency}.}
Finally, an owner also incurs maintenance costs as a result of sharing. We denote these costs by $cs_i$ as in the case of self-usage.

Combining the three components of an owner's cost and reward yields an overall utility for owner $i$ of
\vspace{-.06in}
\begin{equation} \label{eq:utility_owner}
U_{i}(x_{i},s_{i})=f_{i}(x_{i})+p\min\{\frac{D(p)}{S(p)},1\}s_{i}-cx_{i}-cs_{i}.
\end{equation}

Naturally, each owner $i$ determines sharing and self-use levels $s_i$ and $x_i$ by optimizing (\ref{eq:utility_owner}), i.e., choosing $x_i^*(p)$ and $s_i^*(p)$ by solving the following optimization problem:
\begin{eqnarray}
\max_{x_i, s_i} \quad U_i(x_{i},s_{i}), \quad
\text{s.t.}\quad x_i\geq 0, \,\,  s_i\geq 0,\,\,  s_{i}+x_{i}\leq 1. \label{eq:optmize}
\end{eqnarray}
Crucially, the utility function above couples every owner's utility and thus the model yields a \emph{game} once the sharing price $p$ is set by the platform.

It is important to emphasize the asymmetry of the owner and renter models.  This asymmetry captures the fact that product owners often stay for longer time in the sharing platform. Hence, they choose their self-usage and sharing levels based on the long term overall payoff (represented by price times the long term fraction of time they are matched to requests, i.e., $p\min\{\frac{D(p)}{S(p)},1\}$). In contrast, renters can be seen as ``short term'' participants who care mainly  about whether they can get access to the product with reasonable price each time they need. The contrast between short term price-sensitive consumers and long term providers is common in the literature, especially in the context of online sharing \cite{cachon2015role,Banerjee:2015}.  There is empirical support for this distinction in both academia \cite{chen2015peeking} and industry (Uber) \cite{hall2015effects}.

Finally, it is also important to highlight that the owner and renter models are very distinct from typical models in the literature on two-sided markets, e.g., \cite{weyl2010price} \cite{rochet2006two}, since providers can benefit either from self-usage or sharing, and owners and renters determine their actions asymmetrically.

\subsection{Modeling the platform}

The owners and renters discussed above interact through a sharing platform. The platform matches the owners and the renters and sets a price for exchange of services.  Our focus in this paper is not on algorithms for matching, but rather on the pricing decision.

Note that, given a market price charged by the platform, the owners and renters play a game.  Thus, to begin, we need to define the equilibrium concept we consider.  Specifically, denote $X^*=(x_i^*, i\in\mathcal{O})$ and $S^*=(s_i^*, i\in\mathcal{O})$. Then, a state $(X^*, S^*)$  is called a \emph{Nash equilibrium}, if $(x_i^*, s_i^*)$ is an optimal solution to problem (\ref{eq:optmize}) for all $i\in \mathcal{O}$.

Importantly, every price chosen by the platform yields a different game and, therefore, a (potentially) different set of Nash equilibria.  Thus, \emph{the goal of the platform is to choose a price such that (i) there exists a Nash equilibrium (ideally a unique Nash equilibrium) and (ii) the Nash equilibria maximize a desired objective}. Our focus in this paper is on two common objectives: \textbf{revenue maximization} and \textbf{social welfare maximization}.

Revenue maximization and social welfare maximization represent the two dominant regimes under which sharing economy platforms aim to operate.  A platform focused on maximizing short term profits may seek to optimize the revenue obtained at equilibrium, while a platform focused on long-term growth may seek to optimize the social welfare obtained by owners and renters.

More formally, when aiming to maximize social welfare, the platform's objective is to maximize the following aggregate welfare
\begin{equation} \label{def:sw}
W(p)\triangleq\sum_{i\in \mathcal{O}}U_i(x_i^*,s_i^*)+\sum_{k\in \mathcal{R}}U_k(y_k^*),
\end{equation}
where $x_i^*$, $s_i^*$, and $y_k^*$ are the optimal actions by users under the price $p$. Notice that the welfare is defined over all possible \emph{uniform price policies}. An optimal social welfare policy maximizes $W(p)$ by choosing the optimal $p$.

In contrast, when aiming to maximize revenue, the platform tries to  maximize
\begin{eqnarray} \label{def:volume}
R(p)\triangleq p\min\{D(p), S(p)\}.
\end{eqnarray}
The adoption of this objective is motivated by the fact that in many sharing systems, the platform obtains a commission from each successful transaction, e.g., Uber charges its drivers 20\% commission fee \cite{uberfee}. Thus, maximizing the total transaction volume is equivalent to maximizing the platform's revenue.

In both cases, to ensure non-trivial marketplace governed by the platform, we make the following assumptions.

\vspace{-0.1in}
\begin{assumption}[The market is profitable]\label{assumption:pdp}
There exists at least one price $p$ such that $pD(p)\geq c$.
\end{assumption}
\vspace{-0.1in}
This assumption is not restrictive and is used only to ensure that there are owners interested in sharing.  Specifically, if no such price exists, it means that the term $p\min\{\frac{D(p)}{S(p)},1\}s_{i}- cs_{i}<0$ in (\ref{eq:utility_owner}).

\vspace{-0.1in}
\begin{assumption}[No owner has a monopoly]\label{assumption:utility-function}
For any owner $i\in\mathcal{O}$, $\partial_{-}f_i(1)< p_o$, where $\partial_{-}f_i(1)$ is the left derivative of $f_i(x_i)$ at $x_i=1$ and $D(p_o)=1$.
\end{assumption}
\vspace{-0.1in}

It is not immediately clear from the statement, but this assumption ensures that, when $S(p)>0$, at least two owners will be sharing in the platform. More specifically, Assumption \ref{assumption:utility-function} essentially requires that product owners' per-unit self-use utilities are upper bounded by some $p_{o}$, such that they will at least start to share their products when price is high enough that only $1$ demand is left in the market.  A formal connection between the technical statement and the lack of a monopoly owner is given in \cite{technical}.

\section{Platform Pricing}\label{sec:main}
The first contribution of this paper is a set of analytic results describing how a sharing platform can design prices that maximize revenue and social welfare.  To obtain these results we first characterize the equilibria among owners and renters for any fixed market price set by the platform (Section \ref{subsect:characterization}).  These results are the building block that allow the characterization of the prices that maximize social welfare and revenue of the platform (Section \ref{subsec:welfare and revenue}).  Then, using these characterizations, we contrast the prices and the resulting efficiency of the two approaches for platform pricing.

Throughout, the key technical challenge is the coupling created by the inclusion of $S(p)$ in the utility functions.  This coupling adds complexity to the arguments and so all detailed proofs are deferred to the appendices in the extended technical report \cite{technical}.

\subsection{Characterization of equilibria}\label{subsect:characterization}

In order to study optimal pricing, we must first characterize the equilibria among owners and renters given a fixed market price.  In this subsection we establish structural properties of the sharing behavior, including the existence of an equilibrium and the monotonicity of supply and demand.

To begin, define a quantity $p_{upper}$, which will be shown to be an upper bound on market prices, as follows:
\begin{eqnarray}
p_{upper}\triangleq \sup\{p \,|\, p D(p)=c\}.
\end{eqnarray}
Note that $p_{upper}$ always exists due to Assumption \ref{assumption:pdp}. Further,  $pD(p)$ is continuous (see the extended version \cite{technical}).

Using $p_{upper}$, we can characterize the market as operating in one of four regimes, depending on the market price.

\begin{theorem}\label{thm:ne_exist}
For any given market price $p$, there exists a Nash equilibrium. Moreover, equilibrium is unique if $f_i(\cdot)$ is differentiable.
In addition, equilibrium behavior of the market falls into one of the following four regimes:
\begin{enumerate}[(i)]
  \item $p\in [0, p_c)$: $S(p)<D(p)$ and $S(p)$ is non-decreasing
  \item $p=p_c$: $S(p) = D(p)$
  \item $p\in(p_c,p_{upper}]$: $S(p)\geq D(p)$
  \item $p> p_{upper}$: $S(p)=0$
\end{enumerate}
\end{theorem}

Some important remarks about  Theorem \ref{thm:ne_exist} follow.  First, existence and uniqueness of market equilibrium under any fixed sharing price is critical for analysis of the market.

Second, the price $p_c$ is the lowest market clearing price, i.e.,  $p_c=\min\{p\,|\,D(p)=S(p)\}$. Establishing this result is not straightforward due to the discontinuity of the derivative of the coupling term $\min\{\frac{D(p)}{S(p)}, 1\}$, which can potentially generate discontinuous points in the $S(p)$ function as owners may find sharing more valuable if there is a slight change in price. Thus, the proposition is proven by carefully analyzing the $S(p)$ function around  $p_c$.

Third, as $p$ increases, supply will continue to increase as long as it is less than demand. However, once supply exceeds demand, it remains larger than demand until, when the price is higher than some threshold price $p_{upper}$, supply drops to $0$ since the market is not profitable.

Finally, note that regimes (ii) and (iii) in Theorem \ref{thm:ne_exist} are  the most practically relevant regimes since sharing demand can all be fulfilled. In this regime, we can additionally prove the desirable property that the platform's goal of revenue maximization is aligned with boosting sharing supply, i.e., a price for higher revenue leads to higher supply.

\begin{theorem}\label{thm:revenue-supply}
For all $p> p_c$ such that $S(p)>D(p)$, supply is higher when revenue is higher, i.e,  $p_{1}D(p_{1})\geq p_{2}D(p_{2})$ implies $S(p_1)\geq S(p_2)$.
\end{theorem}

An important observation about Theorem \ref{thm:revenue-supply} is that  revenue maximization oriented sharing actually encourages more sharing from owners. This is a result that is consistent with what has been observed in the sharing economy literature, e.g., \cite{chen2015dynamic}.  The intuition behind this result is that the pursuit of actual trading volume $pD(p)$ is in  owners' interests, and this objective increases their enthusiasm for sharing. To be specific, the ``effective price'' $\frac{pD(p)}{S(p)}$  seen by an owner has $pD(p)$, the transaction volume, as the numerator. Therefore, a higher trading volume magnifies the ``effective price,'' which in turn stimulates supply.

\subsection{Social welfare and revenue maximization}\label{subsec:welfare and revenue}

Given the characterization of equilibria outcomes in the previous subsection, we can now investigate how a sharing platform can design prices to maximize social welfare or revenue.  Recall that maximizing revenue corresponds to a short-term approach aimed at maximizing immediate profit; whereas maximizing social welfare corresponds to a long-term view that focuses on growing participation in the platform rather than on immediate revenue.

Our first result characterizes the platform prices that maximize social welfare and revenue, respectively.

\begin{theorem}\label{thm:pr_psw}
The social welfare maximizing price, $p_{sw}$, and the revenue maximizing price, $p_r$, satisfy the following:
\begin{enumerate}[(i)]
  \item The lowest market clearing price $p_c$ achieves maximal social welfare.
  \item The revenue-maximizing price is no less than the social welfare maximizing  price, i.e.,  $p_r \geq p_{c}$.  Thus, $S(p_{r})\geq D(p_{r})$.
  \item Revenue maximization leads to better quality of service than social welfare maximization, i.e., $\frac{D(p_{sw})}{S(p_{sw})}\geq\frac{D(p_r)}{S(p_{r})}$.
\end{enumerate}
\end{theorem}

There are a number of important remarks to make about the  theorem.  First, note that both $p_{sw}$ and $p_r$ ensure that supply exceeds demand at equilibria. This is important for the health of the platform.

Second, note that Part (i) implies that more sharing from owners does not necessarily imply a higher social welfare.  This is perhaps counterintuitive, however, it is due to the fact that a higher supply at a higher price can lead to more ``idle'' sharing, which leads to lower utility for everyone.

Third, in Part (iii) we use $\frac{D(p)}{S(p)}$ as a measure of ``quality of service'' since the experience of renters improves if there is proportionally more aggregate supply provided by owners via sharing.  It is perhaps surprising that revenue maximization leads to better quality of service, but it is a consequence of the fact that revenue maximization is aligned with incentives for sharing (see Theorem \ref{thm:revenue-supply}).

\begin{figure*}
\centering
\subfigure[The impact of cost ($N_O=100$ and $N_R=300$).] { \label{fig:price_cost}
\includegraphics[width=0.6\columnwidth]{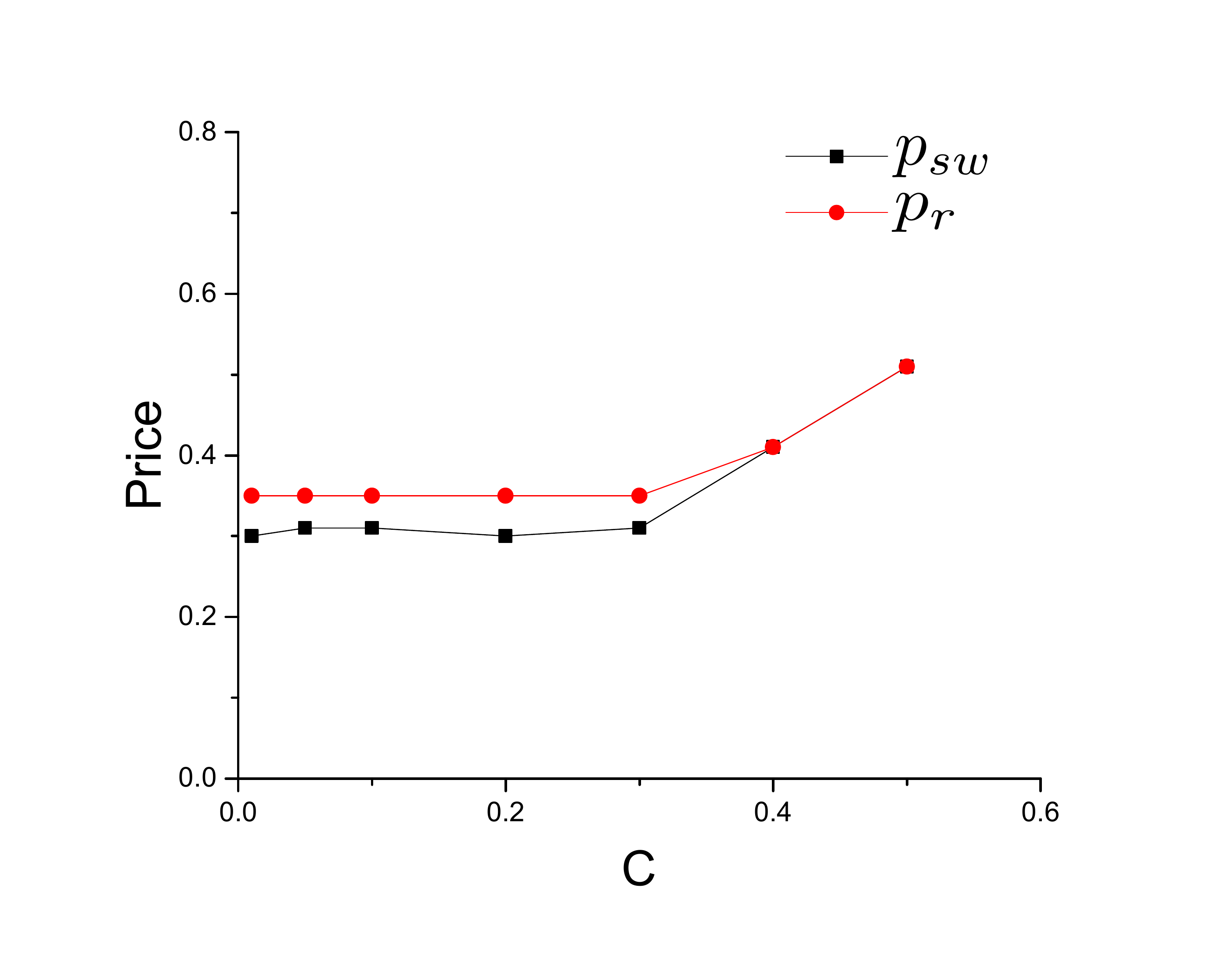}
}
\subfigure[The impact of the number of renters ($N_O=100$ and $c=0.1$).] { \label{fig:price_renter}
\includegraphics[width=0.6\columnwidth]{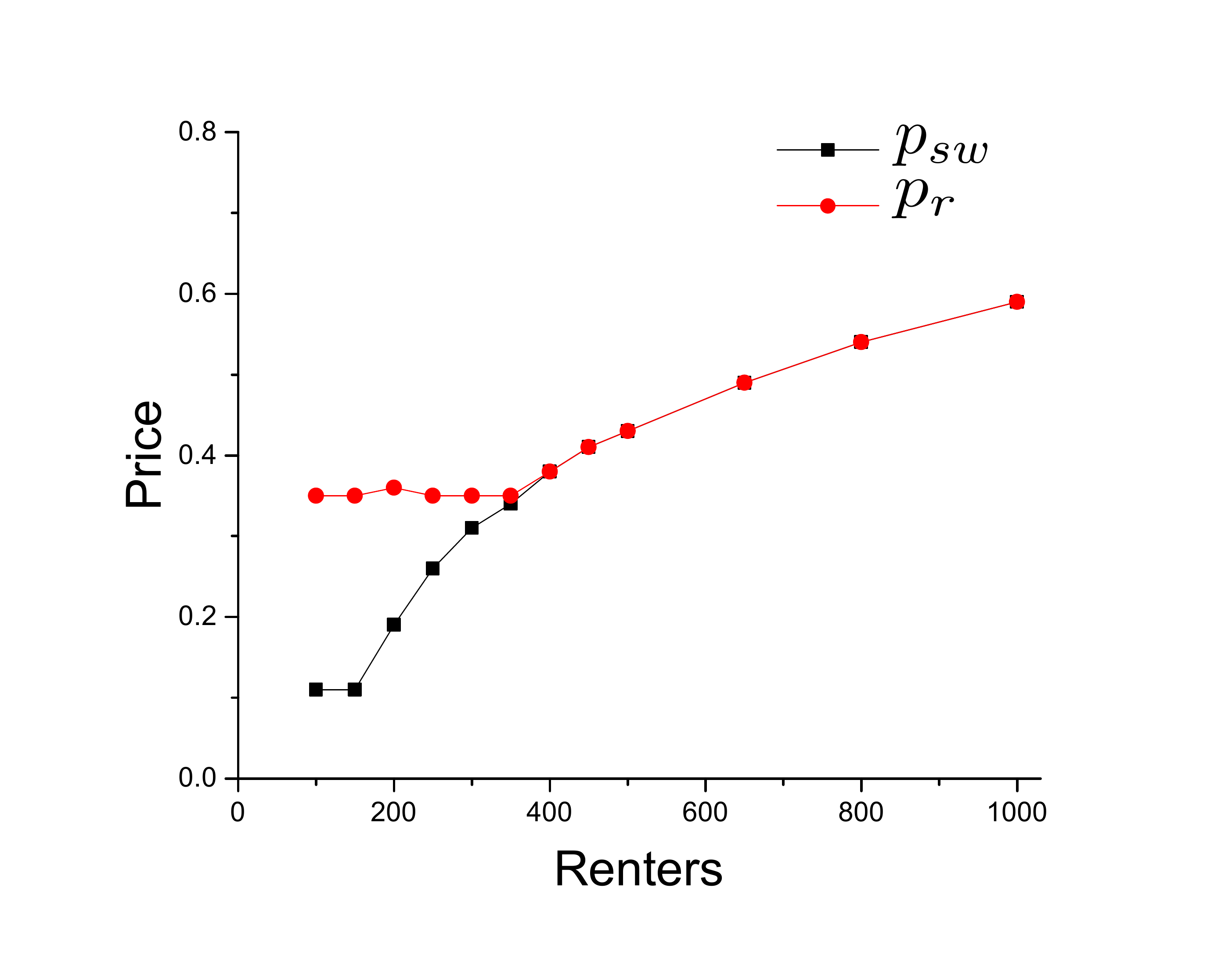}
}
\subfigure[The impact of the number of owners ($N_R=500$ and $c=0.1$).] { \label{fig:price_owner}
\includegraphics[width=0.6\columnwidth]{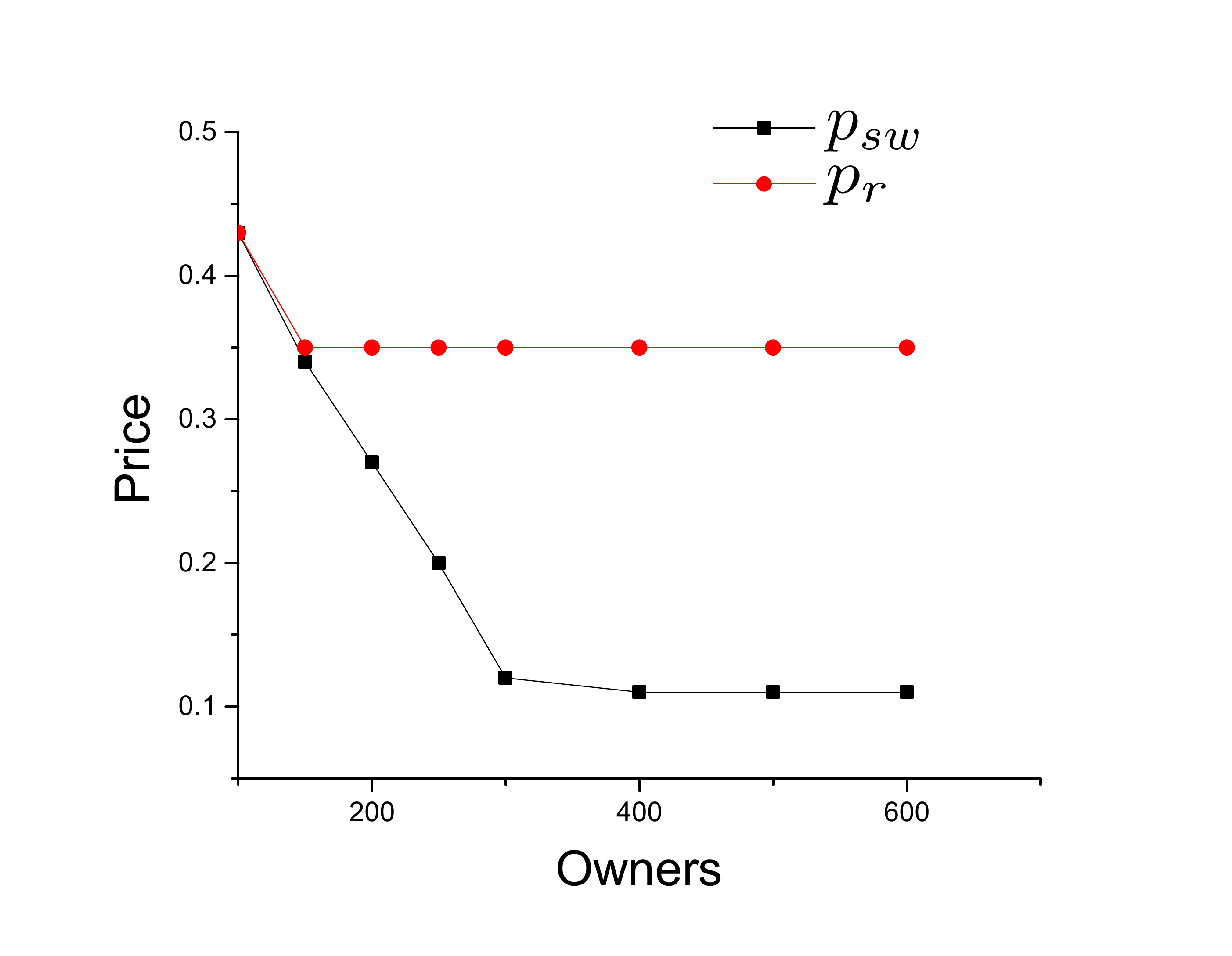}
}
\caption{Comparison of the social welfare maximizing price $p_{sw}$ and the revenue maximizing price $p_r$ under differing costs and resource levels.}
\label{fig}
\end{figure*}

To investigate the relationship between the social welfare and revenue maximizing prices in more detail, we show numerical results in Figure \ref{fig}.  In particular, Figure \ref{fig} shows the relationship of $p_{r}$ and $p_{sw}$ under different costs and different scarcity levels for the products.  The results are shown for quadratic $f$ and $g$.\footnote{ In this simulation, we used quadratic benefit functions of the form $f_i(x_i)=-a_{i}x_{i}^2+b_{i}x_{i}$ (same with $g_k(y_k)$) with $a_{i},b_{i}>0$ for each owner, and $a_i, b_i$ are uniformly  chosen from $(0.1,1.2)$ and $(0,1)$. The numbers of owners and renters vary in different cases.}

Figure \ref{fig:price_cost} highlights that $p_{sw}$ and $p_r$ remain unchanged when costs $c$ are low, but that $p_{sw}$ increases quickly towards $p_{r}$ (eventually matching $p_r$) as $c$ increases.

Additionally, Figure \ref{fig:price_renter} highlights that product scarcity also influences $p_{sw}$ and $p_r$ significantly. When resources become scarce, $p_{sw}$ rapidly increases towards $p_r$.  The intuition for this is that when supply is insufficient, a maximum social welfare policy also needs to guarantee the usage of renters with high utilities, which is the same as what a maximum revenue policy aims to accomplish.  Therefore, when supply is abundant (Figure \ref{fig:price_owner}), $p_{sw}$ is much lower than $p_r$: a maximum social welfare policy wants to fulfill more demand while a maximum revenue policy focuses on high value clients.

This distinction motivates an important question about revenue maximizing pricing: \emph{how much welfare loss does revenue maximizing pricing incur? } Our next theorem addresses this question by providing a ``price of anarchy'' bound.  Recall that $W(p)$ is defined in (\ref{def:sw}) as the aggregate social welfare under price $p$.  Further, we use $s_i^{r}$ to denote the supply of owner $i$ at $p_{r}$.

\begin{theorem} \label{thm:gap}
The social welfare gap between the maximum social welfare policy and a maximum revenue policy is bounded by:
\begin{align}\label{eq:thm3_sum}
0\leq &\,\, W(p_{sw})-W(p_r)    \\
\leq &\,\, p_{r}(D(p_{sw})-D(p_{r}))+p_{r}\frac{D(p_{r})}{S(p_{r})}\sum_{i\in\script{O}: s_i^r\geq s_i^{sw}}(s_i^{r}- s_i^{sw}).\nonumber
\end{align}
In particular, if  $s_i^{r}\geq s_i^{sw}$ for all $i\in\mathcal{O}$, the bounds above become:
\begin{equation}\label{eq:thm3}
  0\leq W(p_{sw})-W(p_r) \leq  p_{r}[D(p_{sw})-D(p_{r})\frac{S(p_{sw})}{S(p_{r})}].
\end{equation}
Moreover, the above bounds are tight.
\end{theorem}

One important feature of the efficiency loss of revenue maximizing pricing is that it can be evaluated by third-party organizations since it only depends on the total demand and supply under prices $p_r$ and $p_{sw}$, not on private utility functions of the owners and renters.

However, the form of the bound in Theorem \ref{thm:gap} does not lend itself to easy interpretation. A more interpretable form can be obtained as follows:
\begin{align}
&W(p_{sw})-W(p_{r}) \nonumber\\
\leq &\,\,  p_{r}[D(p_{sw})-D(p_{r})]+p_{r}\frac{D(p_{r})}{S(p_{r})}[S(p_{r})-S(p_{sw})]. \label{eq:linear_gap}
\end{align}
The first item on the right-hand-side of (\ref{eq:linear_gap}) is an upper bound for the utility loss of renters, since $D(p_{sw})-D(p_{r})$ is the demand decrement from renters at $p_r$, while $p_{r}$ is the upper bound of these renters' usage benefit.
Similarly, the second item is the upper bound of the utility loss of owners since $S(p_{r})-S(p_{sw})$ can been seen as  supply from some new owners starting to share  at $p_{r}$, under risk of meeting no renters,  and $p_{r}\frac{D(p_{r})}{S(p_{r})}$ is the upper bound of their utility loss.

\begin{figure}
\vspace{-.2in}
\centering
\includegraphics[scale=0.22]{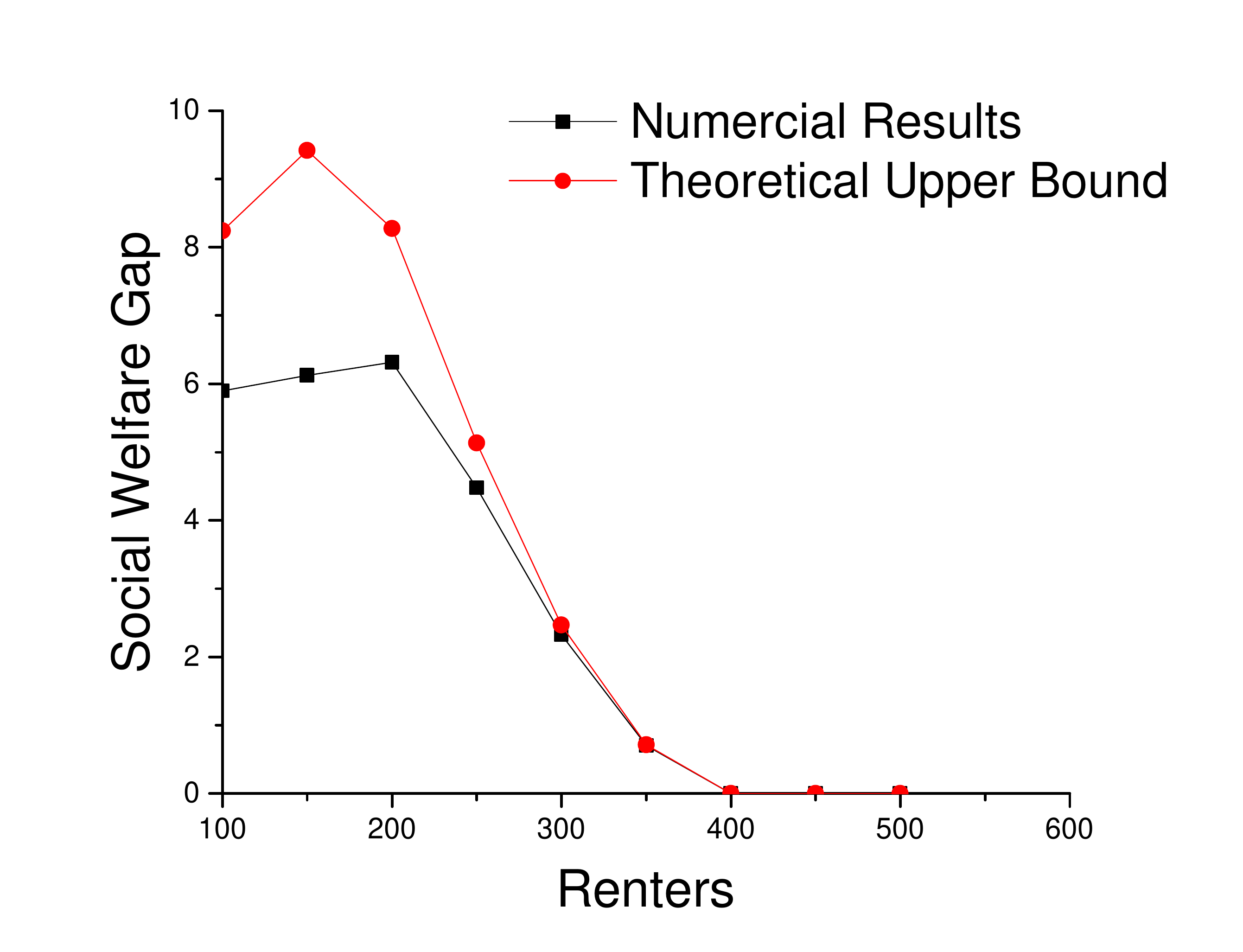}
\vspace{-.2in}
\caption{An illustration of the bound tightness in Theorem \ref{thm:gap}, with same setting used for Figure \ref{fig}. }
\label{fig:supply_demand}
\vspace{-0.2in}
\end{figure}

To obtain  more insight about the bound, Figure \ref{fig:supply_demand} shows numeric results for the case of quadratic $f,g$ and $N_{O}=100$.  Figure \ref{fig:supply_demand} illustrates that, as the number of renters increases, resource becomes scarce and the bound on the social welfare gap decreases to zero. Thus, the revenue maximizing price can be expected to achieve nearly maximal social welfare in platforms where usage is high.

Another important point about the bound on the welfare loss of revenue maximizing pricing is that the bound is tight.

In fact, we show with an example below that the upper bound is  tight for  linear $f,g$.

\begin{example}[Efficiency loss with linear  benefits]

Consider linear usage benefits, i.e.,
\begin{align} \label{eq:linear_usage}
f_i(x_i)=&\alpha_{i}x_{i},\,\,\,\forall i\in \mathcal{O}\nonumber \\
g_k(y_k)=&\alpha_{k}y_{k},\,\,\,\forall k\in \mathcal{R}
\end{align}
Without loss of generality, suppose no users have the same private value. We label and rank all users, including owners and renters, by their private values:
\begin{equation}\label{eq:alpha}
\alpha_{1} > \alpha_{2}  >... > \alpha_{N_{R}+N_{O}}.
\end{equation}
Note that linear utility are a common case of interest in network economics, see \cite{benjaafar2015peer}.

In this case, owner $i$'s best response is to choose either to share his product all the time ($s_i=1$) or to use his product all the time ($x_i=1$), depending on whether his or her private value $\alpha_i$ is lower or higher than the market price.

Similarly, renter $k$ will choose to request  either full use of a product ($y_k=1$) or zero demand ($y_k=0$), depending on his or her private value $\alpha_k$.

We can find that the maximum social welfare price is set such that users (including owners and renters) of  the first $N_O$ highest  $\alpha_{i}$ get full use of the products ($x_i=1$).

This is the  best possible arrangement for maximizing social welfare, since all users with higher private values get access to products. In this case, demand equals supply and no  product is wasted, e.g., no cars are running empty.

Now, we can evaluate the performance of the bound in Theorem \ref{thm:gap}. Consider the concrete example shown in Table \ref{tab:example}, where $N_{R}=N_{O}=3$.

\begin{table}[!htb]
\centering
\begin{tabular}{|c|c|l|} \hline
Ranking of private value &Owner's $\alpha_i$ &Renter's $\alpha_k$\\ \hline
1 & 5 & $4+\delta>4$ \\ \hline
2 & 2 & 1.5 \\ \hline
3 & 1 & 0.5\\ \hline
\end{tabular}
\caption{Tightness of Theorem \ref{thm:gap} for Example 1.}
\end{table}\label{tab:example}

Let $c=0.01$. From the above reasoning, $p_{sw}\in(1.5,2)$, which lets renter $r_1$ use owner $o_3$'s product and  owner $o_1$, $o_2$ use their own products. The transaction volume is  $1\cdot p_{sw}\in(1.5,2)$  and the social welfare is $11+\delta -3c$.
In contrast, $p_r=4+\delta/2$, and is such that only renter $r_1$ can rent a product and owners $o_2$, $o_3$ share their products all the time. The trade volume is  $4+\delta/2$  and  the social welfare is $9+\delta-3c$.

Thus, the social welfare gap between these two policies is $W(p_{sw})-W(p_{r})=2$, and the bound given by Theorem \ref{thm:gap} is $W(p_{sw})-W(p_{r})\leq p_{r}[D(p_{sw})-D(p_{r})\frac{S(p_{sw})}{S(p_{r})}]=2+\frac{\delta}{4}$.
Since $\delta$ can be arbitrarily small, the bound is tight.
\end{example}

\section{Subsidizing Sharing}\label{sec:subsidy}

One may expect that many practical sharing platforms operate in a ``underprovisioned'' or ``congested'' regime where more people will be interested in using the platform to rent than to share.  This is especially true when sharing platforms are getting started.  To handle such situations, most sharing platforms provide subsidies to encourage owners to share, thus increasing supply \cite{ubersubsidy}. These subsidies are crucial to growing supply in order to match demand and our goal in this section is to investigate how a sharing platform can optimize such subsidies.

To begin, recall an important observation from the previous section contrasting social welfare maximization and revenue maximization: the two pricing strategies align in situations where the platform is ``underprovisioned'' or ``congested'', i.e., situations where the number of renters is large compared to the number of owners, e.g., see Figure \ref{fig:price_renter}.  This observation provides a motivation for the use of subsidies: they align because they both use the minimal possible market clearing price, $p_{sw}=p_r=p_c$, which ensures $S(p_c)=D(p_c)$.  In such situations, larger supply would allow the price to be lower and more demand to be served -- yielding (potentially) higher revenue and/or social welfare.

The results in this section characterize the potential improvements from subsidization and identify when subsidies can be valuable.

To begin, define function $V(p)\triangleq p\cdot D(p)$ and let $p_{potential}$ denote the price which maximizes $V(p)$ over $p\in [0,p_{upper}]$. Thus, $V(p_{potential})$ is the potential maximum revenue that the platform can obtain if supply is sufficient. However, it may not possible to achieve this as $S(p_{potential})$ can be smaller than $D(p_{potential})$.

Our first theorem highlights that it is not possible to achieve the maximal potential revenue without subsidization if $p_{potential}$ is lower than the welfare maximizing price.
In this case, the revenue maximizing price aligns with the social welfare maximizing price because dropping below that point would lead to a supply shortage and revenue reduction.

\begin{theorem}\label{thm: potential}
When $p_{potential}\geq p_c$, the platform can extract maximum potential revenue by setting $p_{r}=p_{potential}$. Otherwise, $p_{r}=p_{sw}=p_{c}$.
\end{theorem}

\begin{figure}
\centering
\includegraphics[scale=0.24]{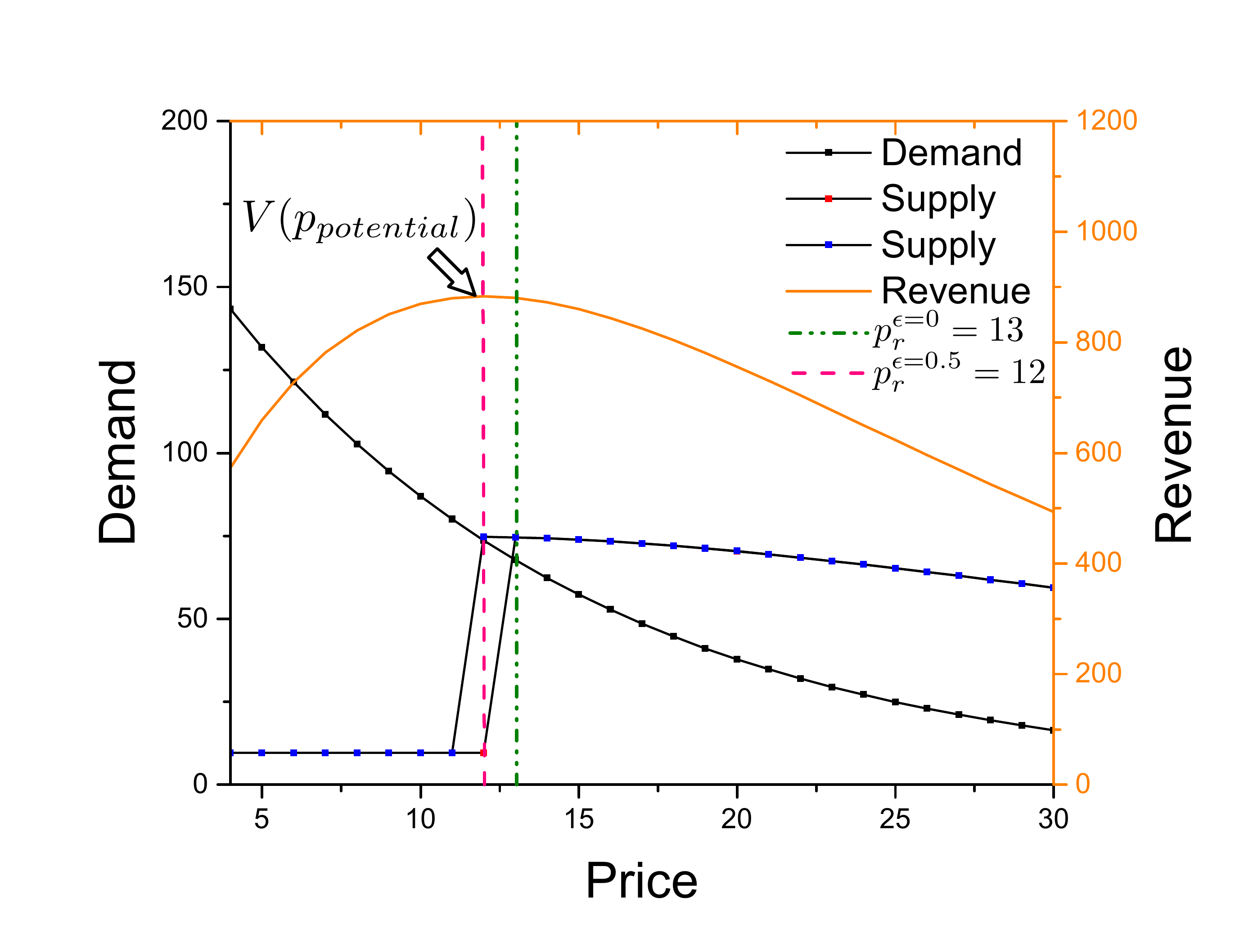}
\caption{Without subsidy, the platform achieves maximum revenue at $p_r=13$ (green line), due to the fact that at $p_{potential}=12$, there is not enough supply.
With subsidy $\epsilon =0.5$, platform can now use $p_{potential}$ to extract maximum potential revenue (red line).
}
\label{fig:epsilon_quadratic}
\end{figure}

The behavior described by Theorem \ref{thm: potential} can be observed in Figure \ref{fig:epsilon_quadratic}, where the red dotted line shows that $p_{potential}=12$ maximizes function $V=pD(p)$. Without subsidy, platform's revenue maximization policy is to set $p_{r}=p_{c}=13$ (the green dotted line), since supply is not sufficient at $p_{potential}$. However, if the platform  subsidizes  owners to boost supply such that supply is higher than demand at $p_{potential}$, than platform can obtain maximum   $V(p_{potential})$.

Figure \ref{fig:epsilon_quadratic} highlights the potential gains from subsidization, but the key question when determining subsidies is if the benefit from subsidization exceeds the cost of the subsidies themselves.
To understand this, we quantify how much additional supply can be obtained by subsidies below.

Specifically, we assume that sharing platform provides an additional subsidy for sharing of $p\epsilon,  \epsilon>0$ per unit sharing. Thus, the effective price seen by an owner becomes $p\cdot (\frac{D(p)}{S(p)}+\epsilon)$.\footnote{Note that this model also applies to the case of enthusiastic users who have an optimistic perception about the fraction of time they will receive customers.}
Therefore, an owner's utility function becomes:
\begin{equation*}
U^{\epsilon}_{i}(x_{i},s_{i}) = f_{i}(x_{i})+p(\min\{\frac{D(p)}{S(p)},1\}+\epsilon)s_{i} -cx_{i}-cs_{i}.
\end{equation*}

Our next  theorems characterize the supply $S^{\epsilon}(p)$ under subsidy factor $\epsilon>0$, compared to the original  $S(p)$ ($\epsilon =0$).  The first result highlights that subsidization necessarily increases aggregate supply.

\begin{theorem}\label{thm:subsidy-0}
For any $\epsilon_2>\epsilon_1\geq0$, we have $S^{\epsilon_2}(p)\geq S^{\epsilon_1}(p)$. In particular, $S^{\epsilon}(p)\geq S^0(p) $, where $S^0(p)$ is the original supply with $\epsilon=0$.
\end{theorem}

The second result highlights that there are three regimes for subsidization, of which the first is the most relevant for practical situations.

\begin{theorem}\label{thm:subsidy}
The impact of subsidization can be categorized into three regimes:
\begin{enumerate}
  \item Small subsidies: For $\epsilon$ such that $\epsilon \leq \frac{D(p)}{S^0(p)}(1-\frac{s_i^0}{S^0(p)})$,$\forall i\in\mathcal{O}$,  we have
  \begin{equation}\label{eq:sqrt_epsilon}
    S^\epsilon(p)\leq F(\epsilon)= \frac{D(p)+\sqrt{M^2+4D(p)s_i^0\epsilon}}
 {2[\frac{D(p)}{S^0(p)}(1-\frac{s_i^0}{S^0(p)})-\epsilon]},
  \end{equation}
  where $M=D(p)(1-\frac{2s_i^0}{S^0(p)})$. Also, $F(0)=S^0(p)$.

  \item Medium subsidies: For $\epsilon$ such that $ \frac{D(p)}{S^0(p)}(1-\frac{s_i^0}{S^0(p)}) < \epsilon < \frac{1}{p}(\partial_{+}f_i(0)-\frac{pD(p)S_{-i}^{\epsilon}}{(S_{-i}^{\epsilon}+1)^2})$, we have that  $x_i^\epsilon(p)+s_i^\epsilon(p)=1$ and $s_i^\epsilon < 1$ for all $i\in \mathcal{O}$.

  \item Large subsidies: For $\epsilon$  such that $\epsilon\geq \frac{1}{p}(\partial_{+}f_i(0)-\frac{pD(p)S_{-i}^{\epsilon}}{(S_{-i}^{\epsilon}+1)^2})$, for all $i\in\mathcal{O}$,  we have that $s_i^\epsilon =1$, i.e., $S^{\epsilon}(p)=N_O$.

\end{enumerate}
\end{theorem}

The above theorems together highlight that subsidies necessarily increase the sharing supply from product owners. Without subsidy, when there is a supply shortage at $p_{potential}$, the platform's maximum revenue policy is to set price to $p_c$.
However, if the platform chooses a proper subsidy $\epsilon$ and a price $p^\epsilon$ then it can achieve improved revenue. The following theorem characterizes when this is possible.

\begin{theorem}\label{theorem:subsidy}
There exists a subsidy $\epsilon$ that increases revenue whenever
\[V^\epsilon(p)=p^\epsilon \min\{D(p^\epsilon),S^{\epsilon}(p^\epsilon)\}-\epsilon S^{\epsilon}(p^\epsilon)\geq p_{c}D(p_{c}).\]
\end{theorem}

Note that Theorem \ref{theorem:subsidy} also shows that an appropriate subsidy $\epsilon$ can be found by optimizing $V^{\epsilon}(p)$.
Concretely, Figure \ref{fig:epsilon_quadratic} shows that by introducing  $\epsilon=0.5$, the platform can  close the supply gap at $p_{potential}=12$ by shifting the market clearing price from $p_c=13$ to $p_{potential}=12$.

Beyond the example in Figure \ref{fig:epsilon_quadratic}, we also derive the impact of subsidies analytically under quadratic usage benefits in the following example, showing substantial increase in revenue.

\begin{example}[Subsidies for quadratic usage benefits]

Suppose there are $N_{O}=100$ owners and $N_{R}=150$ renters in the platform.
Let all renters' utility functions be $g_k(y_k)=3y_k-y_k^2$,  all owners' utility functions be $f_i(x_i)=4x_i-x_i^2$ and $c=0$.
In this case, total demand is given by $D(p)=N_{R}\cdot(3-p)/2=225-75p$ by solving (\ref{eq:optmize-renter}). Optimizing $V(p)=pD(p)$ gives  $p_{potential}=1.5$.

To derive the market clearing price $p_c$, we first calculate each owner's supply at $p_c$:
\begin{equation*}
  s_i^*(p_{c})=\argmax{s_i}   U_{i}(x_{i},s_{i})=f_{i}(x_{i})+ p_{c}s_{i} = (\frac{p_c-2}{2})_+
\end{equation*}
Here we have used $s_{i}+x_{i}=1$ at $p_c$ since $c=0$. 
This gives $S(p_c)=\sum_{i\in\mathcal{O}}s_i^*(p_{c})=N_{O}\cdot(p_c-2)/2=50p_c-100$.
By solving the following equation
\begin{equation*}
  D(p_{c})=S(p_c)
\end{equation*}
we have $p_{c}=2.6$, which gives $D(p_c)=S(p_c)=30$.
From the above, we see that at $p_{potential}=1.5$, $S(p_{potential})=0<D(p_{potential})=112.5$.
By Theorem \ref{thm:pr_psw} and Theorem \ref{thm: potential},
we have that $p_c=p_r=p_{sw}=2.6$ and platform's revenue is $V_0=p_cD(p_c)=78$.

Now, consider providing a subsidy $\epsilon$. Denote $p_{c}^{\epsilon}$ as the market clearing price under subsidy $\epsilon$.
We have $D(p)=225-75p$ and $S^\epsilon(p)=50(p+\epsilon)-100$ for all $p\leq p_{c}^{\epsilon}$.
By solving $D(p_{c}^{\epsilon})=S(p_{c}^{\epsilon})$ we have:
\begin{equation}\label{eq:pc_subsidy}
  p_{c}^{\epsilon}=(13-2\epsilon)/5.
\end{equation}

Now consider the case when the platform use $p^{\epsilon}=p_{c}^{\epsilon}$,  so that the platform can avoid over-subsidizing owners when supply is higher than demand.
Hence, the net revenue  is given by:
\begin{equation*}
  \max_{\epsilon} \quad V^{\epsilon}=(p_{c}^\epsilon-\epsilon)S^\epsilon(p_c^\epsilon)
\end{equation*}
Plugging (\ref{eq:pc_subsidy}) into the above objective we can find that the optimal subsidy will be  $\epsilon=\frac{3}{7}$.

In this case, we show that by setting subsidy $\epsilon=\frac{3}{7}$, $p_{r}=p_{c}^{\epsilon}=2.43$. In which case we have $S^\epsilon(p_{r}^{\epsilon})=D(p_{r}^{\epsilon})=42.75$, and the transaction volume is $pD(p)=103.9$ and  the platform obtains a higher net revenue $V^{\epsilon}=85>V_0$.
\end{example}

\section{Case Study: Didi Chuxing}\label{sec:simulation}
\begin{figure*} [!ht]\centering
\subfigure[Didi data.] 
{\label{fig:role_of_cost_real}
\includegraphics[width=0.60\columnwidth]{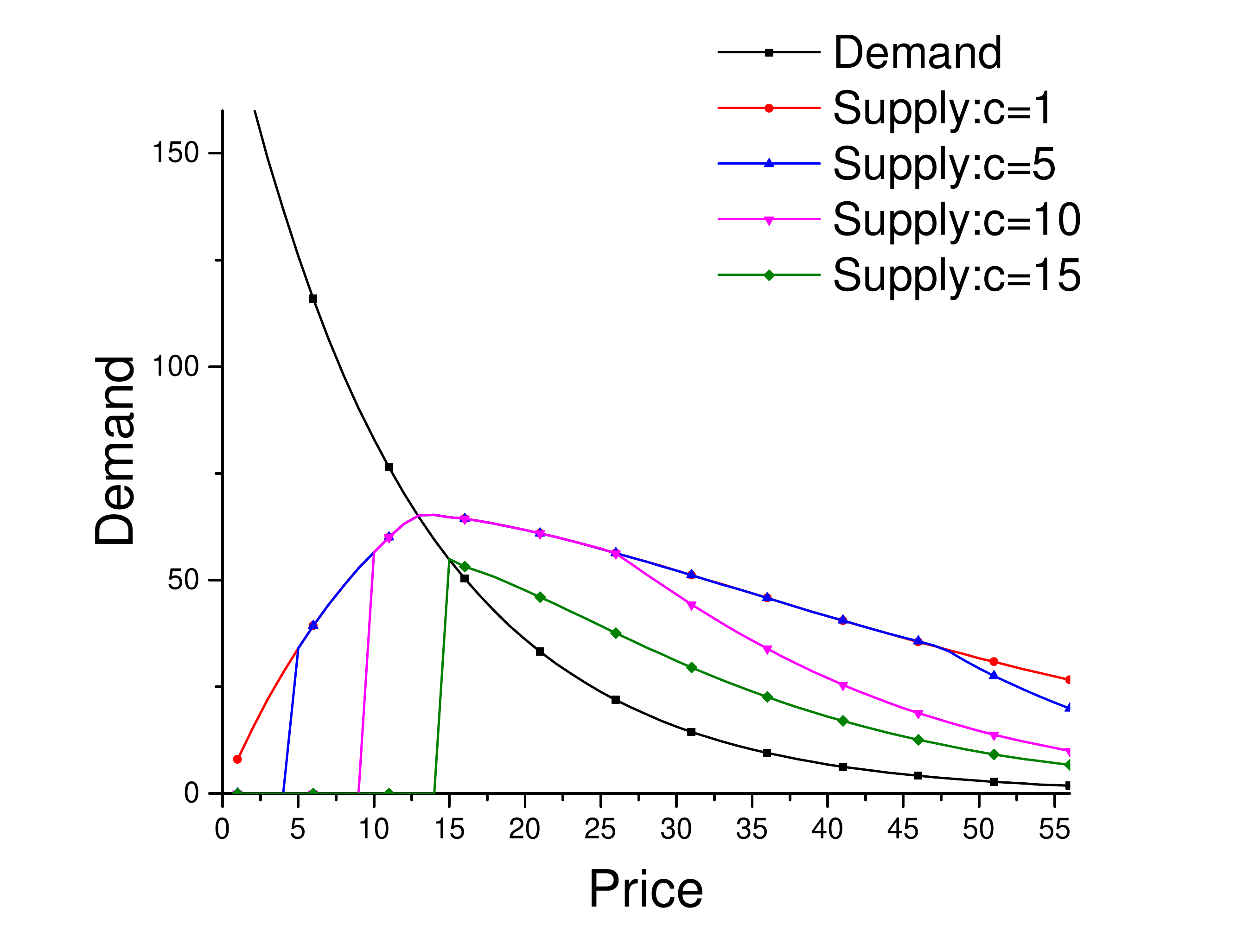}
}
\subfigure[Quadratic model.] 
{ \label{fig:role_of_cost}
\includegraphics[width=0.60\columnwidth]{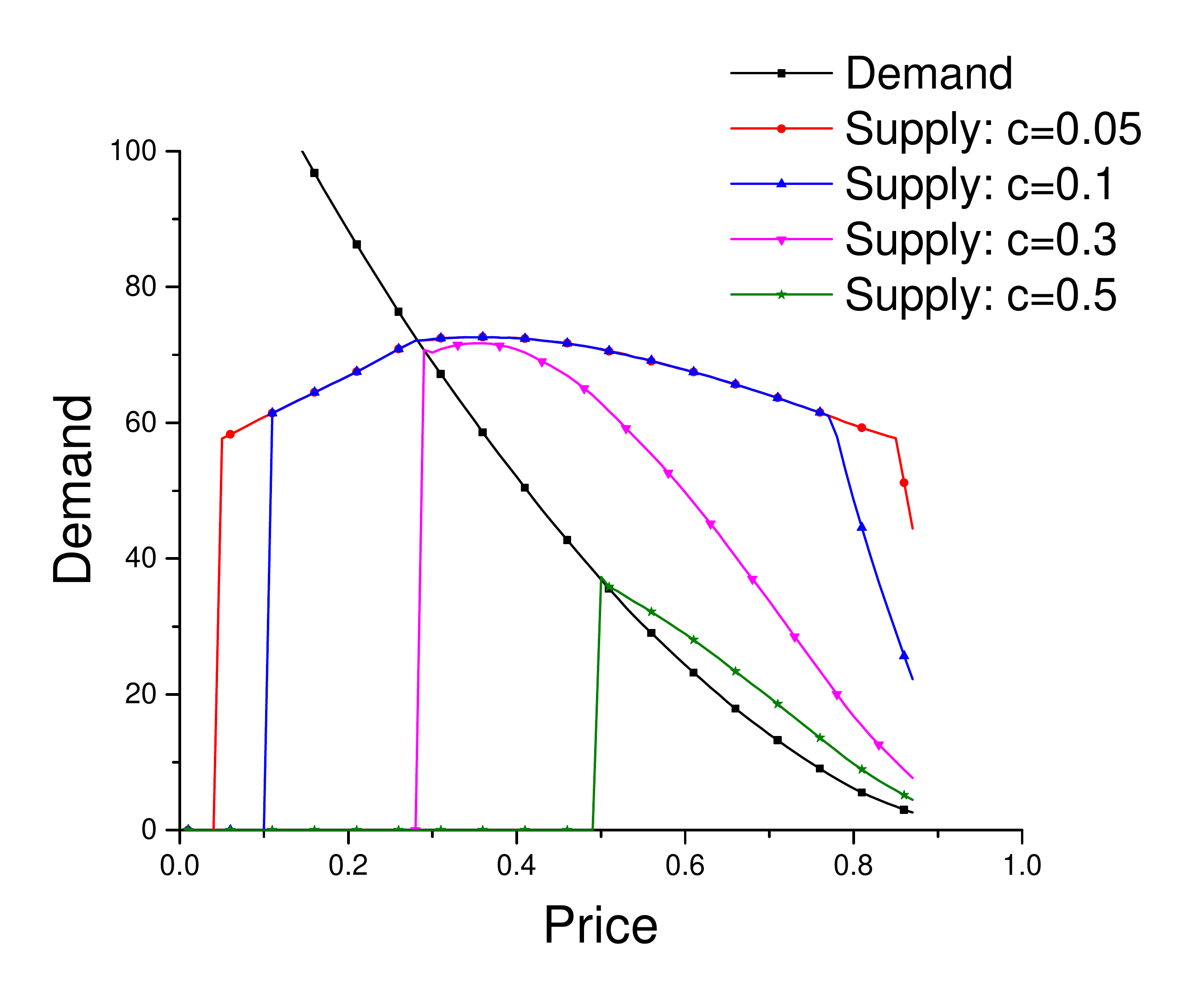}
}
\subfigure[The effect of subsidies.] 
{\label{fig:epsilon}
\includegraphics[width=0.60\columnwidth]{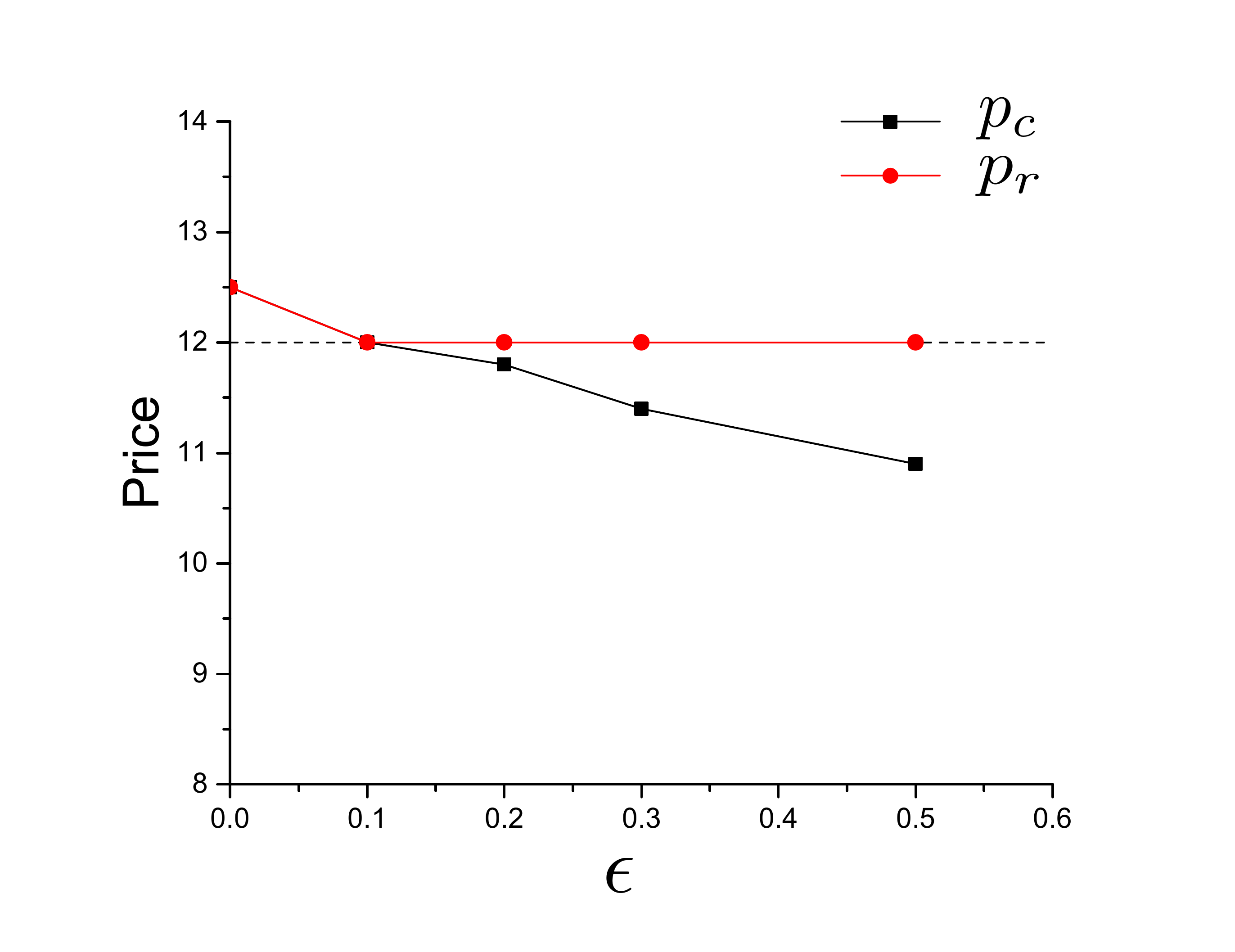}
}
\caption{Role of cost and subsidies.}
\end{figure*}

We consider the case of ridesharing in this section in order to ground the analytic work described above.  We have obtained a dataset from Didi Chuxing, the largest ride-sharing platform in China -- Didi Chuxing sees $10$ million daily rides \cite{didi_intro}. The dataset we have obtained includes transaction records for the first three weeks of January 2016, in a Chinese city. Each record consists of driver ID, passenger ID, starting district ID, destination district ID, and fee for a ride. For the experiments described below we use data from Jan $8$th, which includes $395,938$ transaction records, though results are consistent for other days.

\subsection{Experimental setup}\label{subsec:real}
\begin{figure}[!h]
\centering
\includegraphics[height=1.55in, width=2.2in]{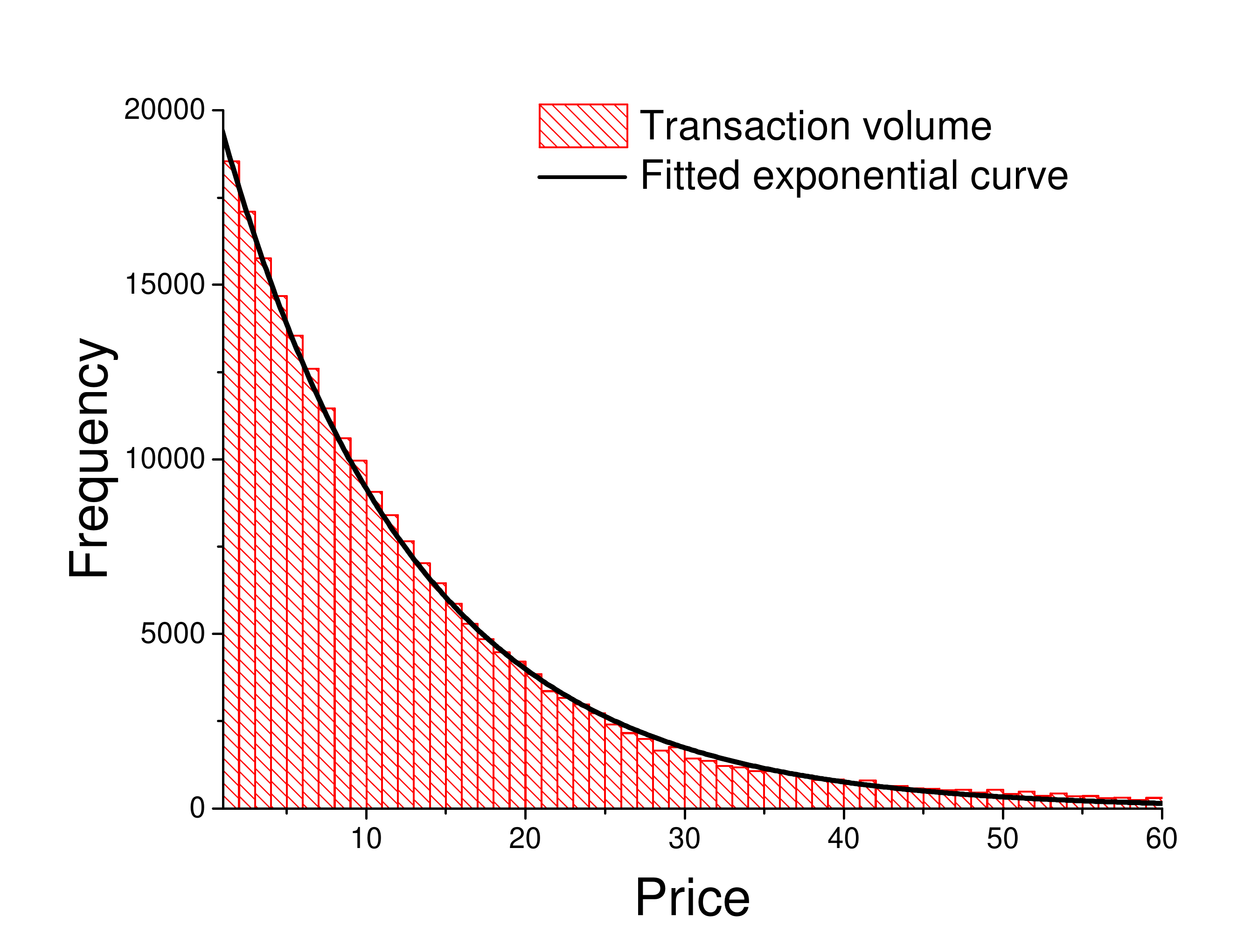}
\caption{Transaction volume from Didi data.}
\label{fig:data}
\end{figure}

In order to fit our model to the data, we use the number of transactions at different prices to represent renters' demands under different prices. This allows us to fit an aggregate demand curve.  We use an exponential function of the form
$D(p)=\alpha e^{-\beta p},$
where $\alpha=19190, \beta=0.0832$. This achieves an $r^2$ value of $0.9991$.  The fit is illustrated in Figure \ref{fig:data}.

The exponential form for the aggregate demand curve corresponds to each renter's $g_k(y_k)$ having following form:
\begin{equation}\label{eq:experiment_utility}
  g_k(y_k)=\frac{1}{\beta}(y_k+y_k\log(\frac{\alpha}{N_R})-y_k\log y_k).
\end{equation}
We also assume that product owners have the same form for their usage benefit.
Finally, since DiDi has a minimum charge on each ride (at $10$ RMB), we reset the zero price point to the minimum charge fee.


\subsection{Experimental results}

Using the dataset from Didi, we study the distinction between social welfare and revenue maximizing prices, the impact of cost, and the role of subsidies in practical settings.

\textbf{Welfare loss.} The first question we ask is ``how do social welfare maximizing prices and revenue maximizing prices differ in practical situations?"  To answer this question, we examine settings when supply is less  than demand, i.e., from  $N_O=100$ to $1100$ while $N_R=1919$.  These ``congested'' scenarios align with the load experienced by Didi.

\begin{figure}[!h]
\centering
\includegraphics[height=1.75in, width=2.4in]{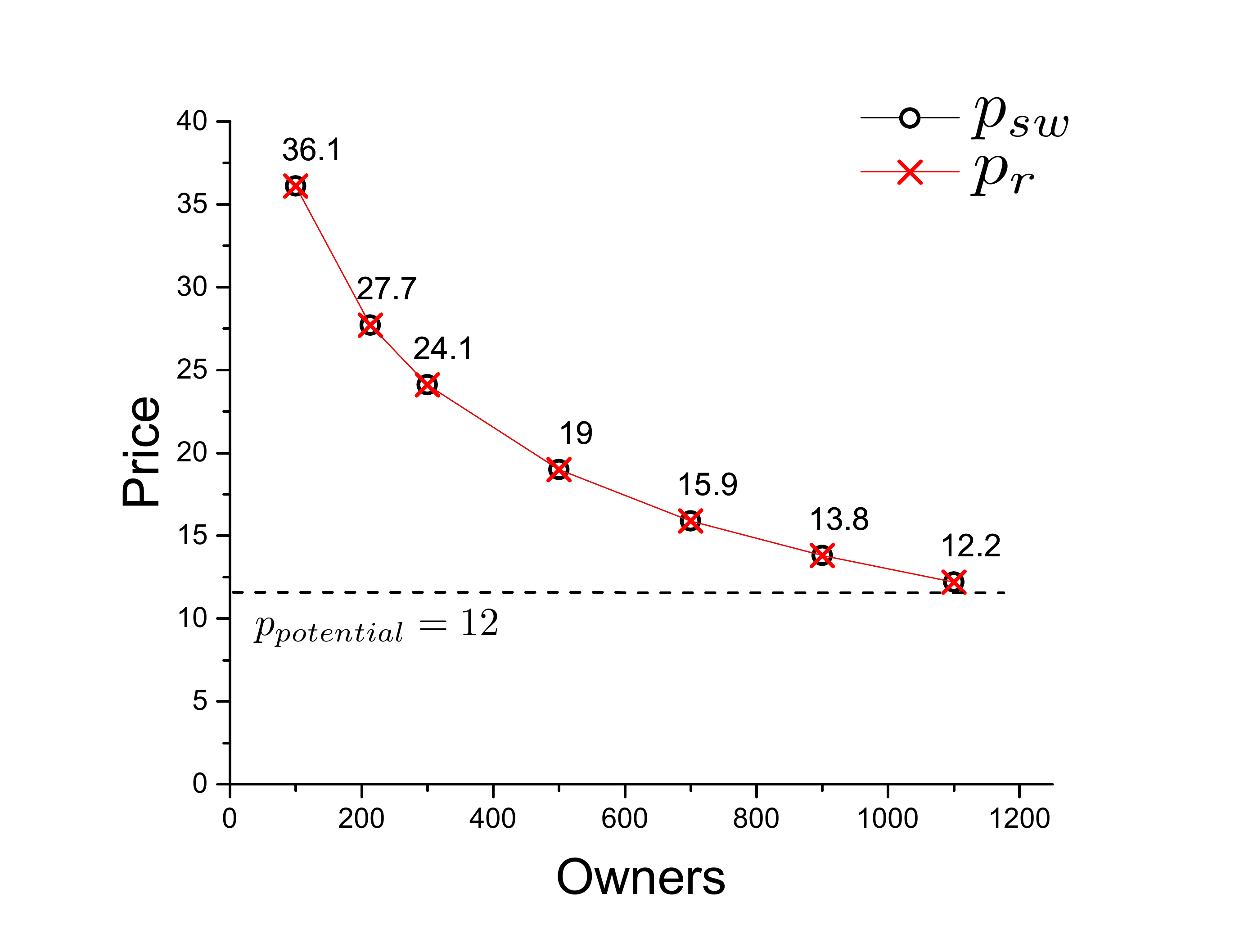}
\caption{Under real world exponential form demand, supply is in shortage and hence $p_{r}=p_{sw}$.}
\label{fig:real_price}
\end{figure}

The results are summarized in Figure \ref{fig:real_price}, which highlights that,  when supply is not abundant, the platform actually achieves maximum social welfare and maximum revenue simultaneously when it sets the price to maximize revenue.
Note that, given the parameterized model of demand, we can calculate the price that produces the maximum potential revenue:
$ p_{potential}=\text{argmax}_{\{p\}} \,\,  pD(p)=12.$

The alignment of revenue and welfare in Figure \ref{fig:real_price} is consistent with the results in Theorem \ref{thm: potential}: since $p_{c}=p_{sw}>p_{potential}$ due to insufficient supply, the revenue maximizing price will be $p_r=p_{sw}$.

\textbf{The role of cost.} Next, we investigate how costs impact platform prices and outcomes.  Recall that the cost $c$ corresponds to the maintenance costs for the product.  Importantly, it also captures any supply-side regulation through taxation on usage, e.g., a gas tax.   The results are summarized in Figure \ref{fig:role_of_cost_real} ($N_O=100$, $N_R=191$), which shows the demand and supply under different costs, and Figure \ref{fig:role_of_cost} ($N_O=100$, $N_R=300$), which shows the results based on  quadratic usage benefit function for comparison.

In these two figures, we  observe that the cost $c$ is an important factor in determining the lowest price to enable sharing, i.e., there will be no sharing at $p<c$. More importantly, the figures highlight that higher costs significantly  suppress sharing. This is intuitive, as drivers choose their sharing level more conservatively if they have to pay more maintenance costs during the times when they are sharing.   This highlights that \emph{a properly chosen cost $c$ can help  reduce redundant supply while maintaining an acceptable QoS}. This is crucial to many social issues such as greenhouse gas emission, road congestion, and it is implementable through, e.g.,  gas taxes.

Another result that is highlighted in Figure \ref{fig:role_of_cost} is that, when $c$ is not high enough, the market clearing price $p_c$ remains the same. This implies that the platform's revenue does not change with costs when cost is at low-to-moderate levels. That is to say,  if usage taxes are used to regulate a platform, product owners will incur this cost and the platform's revenue is likely to not be impacted.

\textbf{The Role of Subsidies.} Finally, we investigate the impact of subsidies in practical settings.  Figure \ref{fig:real_price} highlights that $p_{r}$ differs from the price that optimizes the potential revenue $p_{potential}$.  This is because the platform does not attract enough sharing. Hence, the platform can not extract the maximum potential  revenue since $p_{r}=p_{c}>p_{potential}$ in this case.
As illustrated in Section \ref{sec:subsidy}, subsidizing  owners for sharing is beneficial in such situations.  Figure \ref{fig:epsilon} shows the impact of subsidizing  owners in the Didi data set, with $N_{R}=1919$, $N_O=1100$.
In this case, only a small subsidy ($\epsilon=0.1)$ is needed to steer the price toward $p_{potential}=12$.

\section{Concluding Remarks}\label{sec:conclusion}
In this paper we study the design of prices and subsidies within sharing platforms.  We extend traditional models of two-sided markets to include the opportunity for a tradeoff between using a resource directly or sharing it through a platform.  This extension adds considerable technical complexity.  However, we are able to prove that a unique equilibrium exists regardless of the prices imposed by the sharing platform.  Further, we provide results characterizing the revenue maximizing price and the social welfare maximizing price.  These results allow us to bound the efficiency loss under revenue maximizing pricing.  Additionally, these results highlight that the revenue maximizing price may be constrained due to supply shortages, and so subsidies for sharing are crucial in order to reach the maximal potential revenue.  Finally, we provide results that allow optimization of the size of the subsidy to maximize the tradeoff between increased revenue and the cost of the subsidies themselves.  Our results are grounded by an exploration of data from Didi, the largest ridesharing platform in China.

The results in this paper provide interesting insights about prices and subsidies in sharing platforms, but they also leave many questions unanswered.  For example, we have considered a static pricing setting, and dynamic pricing is common in sharing platforms.  Recent work has made some progress in understanding dynamic pricing in the context of ridesharing, but looking at dynamic pricing more generally in two-sided sharing platforms remains a difficult open problem.  Additionally, our work separates pricing from matching in the sharing platform.  We do not model how matches are performed, we only assume that they are ``fair'' in the long-run.  Incorporating constraints on matching, and understanding how these impact prices, is a challenging and important direction for future work.

\bibliographystyle{abbrv}
\bibliography{acmsmall-market-bibfile}

\section{appendix}
\subsection{Proof of Theorem \ref{thm:ne_exist}}
\begin{proof}
We prove the existence of Nash Equilibrium first, and then prove the characterization of supply and demand in four regimes of $p$.

\noindent \textbf{Proof of Existence}

Let $\mathbf{S}=(s_1, s_2, ...s_{N_o})$, with $s_i$ being owner $i$'s share level, and $\mathbf{X}=(x_1, x_2..., x_{N_O+N_R})$, with $x_i$ being owner $i$'s  self-usage level.

First we notice that an owner $i$'s strategy space is $(x_i, s_i)\in\{x_i\geq0, s_i\geq0, x_i+s_i\leq1\}$ and a renter's action space is $y_k\in[0,1]$, both of which are compact convex sets.
%
The total demand $D(p)$ is fixed as long as $p$ is set, since renters' demand only depend on their utilities and the  market price.
Also note that the game only involves owners, i.e., consider equation (\ref{eq:utility_owner}).
%
Since the total supply $S(p)=\sum_is_i$,  owner $i$'s utility function is continuous in $\{s_1, s_2, ..., s_{i-1}, s_{i+1}, ..., s_{N_O}\}$ and $\mathbf{X}$.

Depending on $D(p)$ and others owners'  total supply $\sum_{j\neq i}s_j$, there are three cases:
\begin{itemize}
\item $D(p)<\sum_{j\neq i}s_j$: the utility function for $i$ becomes $U_{i}(x_{i},s_{i})=f_{i}(x_{i})+p\frac{D(p)}{S(p)}s_{i}-cx_{i}-cs_{i}$ and we know that $p\frac{D(p)}{S(p)}s_{i}=p\frac{D(p)}{\sum_{j\neq i}s_{j}+s_{i}}s_{i}$ is concave in $s_i$.
\item $D(p)>\sum_{j\neq i}s_j+1$: the utility function for $i$ becomes $U_{i}(x_{i})=f_{i}(x_{i})+ps_{i}-cx_{i}-cs_{i}$, which is also concave in $s_i$.
\item $\sum_{j\neq i}s_j<D(p)<\sum_{j\neq i}s_j+1$, the utility function for $i$ becomes
\begin{equation*}
\hspace{-.1in}U_{i}(x_{i},s_{i})=\begin{cases}
f_{i}(x_{i})+p\frac{D(p)}{S(p)}s_{i}-cx_{i}-cs_{i} & \text{if}\,\, D(p)\leq S(p) \\
f_{i}(x_{i})+ps_{i}-cx_{i}-cs_{i} & \text{else}
\end{cases}
\end{equation*}
This function is continuous. It is  concave in $s_i$ since (a) each piece is concave in $s_i$ and (b) at the intersection point of these two pieces, the left derivative $p-c$ is greater than the right derivative $p\frac{D(p)(S(p)-s_i)}{S(p)^2}-c=p\frac{S(p)-s_i}{S(p)}-c$.
\end{itemize}
Hence, owner $i$'s utility function in all three cases above is  continuous and concave to its own strategy $(x_{i}, s_{i})$.

According to \cite{glicksberg1952further}, \cite{debreu1952social}, and \cite{fan1952fixed}, a pure strategy Nash equilibrium exists since each user's strategy space is convex and compact, and the utility function of each user is concave in his strategy, and continuous in $\mathbf{S}$ and  $\mathbf{X}$.

\vspace{.1in}
\noindent \textbf{Proof of  Characterization}

Before proving the characterization of supply and demand, we first introduce the  proposition below, which will be proven in following section.
\begin{prop}
The total demand $D(p)$ is continuous and non-increasing in the market  price $p$ and  eventually reaches $0$ when $p$ is high enough, i.e., $p\geq B$ where $B$ is the upper bound of the right derivative of $g_k(\cdot)$ at $y_k=0$. \label{prop:demand}
\end{prop}
By Proposition \ref{prop:demand},  the total demand decreases to $0$ when market price exceeds $B$. Note that $p_{upper}D(p_{upper})=c$.

(a) We first show by contradiction that there exists a point  $p$ such that $D(p)\leq S(p)$ in $[0, p_{upper})$.
Suppose there is no cross point between  $D(p)$ and $S(p)$, then we have $S(p)<D(p)$ for all $p\in[0,p_{upper})$.

Since each owner's benefit function is strictly concave and its derivative at $0$ bounded by $p_{upper}$, we can always find some $p_{\delta}=p_{upper}-\delta$, where $\delta>0$ can be infinitely small, such that $S(p_{\delta})>0$.
Then, owner $i$'s overall utility is given by:
\begin{equation}
U_{i}(x_{i},s_{i})=f_{i}(x_{i})+p_{\delta}s_{i}-cx_{i}-cs_{i}, \label{eq:prop_pdelta}
\end{equation}
Note that in (\ref{eq:prop_pdelta}), as we inrease $p_{\delta}$ towards $p_{upper}$, owners will not lower their supply, i.e., $S(p)$ is non-decreasing in $[0,p_{upper}]$ and $S(p)\geq S(p_{\delta})>0$ in $[p_{\delta},p_{upper})$.
On the other hand,  $D(p)$ continuously decreases to $0$ as $p$ goes to  $p_{upper}$. These two properties guarantee an intersection of $D(p)$ and $S(p)$ and lead to a contradiction.

(b) We now show Part (ii), i.e.,  there exists $p'<p_{upper}$  such that $D(p')=S(p')$.  Thus, $p_c=p'$ here.

Since  $S(p)$ is eventually larger than $D(p)$, there must exist some price $p'$ such that $D(p'_-)>S(p'_-)$ for $p'_-=p'-\delta$ and $D(p_+)<S(p_+)$ for $p'_+=p'+\delta$, here $\delta\rightarrow0^+$ is  a positive infinitesimal.
We only need to show that $D(p')=S(p')$, i.e., $S(p)$ is continuous at the intersection point $p'$.

Suppose the opposite that $S(p'_+)-S(p'_-)=\Delta>0$, where $\Delta$ is some positive constant.  Then, there must exists some owner $i$, such that $s_i^*(p'_+)-s_i^*(p'_-)=\Delta_i>0$, and the effective price at $p'_+$ is $\frac{p'_+D(p'_+)}{S(p'_+)+\Delta}$.
Let $x_i^{+},s_i^{+}$ and $x_i^{-},s_i^{-}$ denote owner $i$'s strategy at prices  $p'_+$ and $p'_-$ in equilibrium, respectively, and denote $S_{-i}^{+} =\sum_{j\in \mathcal{O}, j\neq i}s_j^{+}$. Since $D(p_+)<S(p_+)$,  we have:
\begin{eqnarray}
&& f_i(x_i^+)+\frac{p'_+D(p'_+)}{S_{-i}^{+}+s_i^{+}}s_i^{+}-cx_i^{+}-cs_i^{+}\nonumber\\
&&\qquad \geq  f_i(x_i^-)+p'_+\min\{\frac{D(p'_+)}{S_{-i}^{+}+s_i^{-}},1\}s_i^{-}-cx_i^{-}-cs_i^{-} \nonumber
\end{eqnarray}
and that
\begin{eqnarray}
f_i(x_i^-)+p'_-s_i^{-}-cx_i^{-}-cs_i^{-}\geq f_i(x_i^+)+p'_-s_i^{+} -cx_i^{+}-cs_i^{+}  \nonumber
\end{eqnarray}
Summing up two equations above, we have:
\begin{equation}\label{eq:prop2_contradict}
  (\frac{p'_+D(p'_+)}{S_{-i}^{+}+s_i^{+}}-p'_-)s_i^{+}\geq (p'_+\min\{\frac{D(p'_+)}{S_{-i}^{+}+s_i^{-}},1\}-p'_-)s_i^{-}
\end{equation}
Since $S_{-i}^{+}+s_i^{+}=S^{+}>D^{+}$, we have $\frac{p'_+D(p'_+)}{S_{-i}^{+}+s_i^{+}}-p'_-<0$ by having a small $\delta$. Now consider the right hand side, there are two cases.
\begin{itemize}
  \item If $\frac{D(p'_+)}{S_{-i}^{+}+s_i^{-}}\geq1$, (\ref{eq:prop2_contradict}) becomes $(\frac{p'_+D(p'_+)}{S_{-i}^{+}+s_i^{+}}-p'_-)s_i^{+}\geq 0$, which means $s_i^{+}\leq 0$ and gives a contradiction.
  \item If $\frac{D(p'_+)}{S_{-i}^{+}+s_i^{-}}<1$,  we have:
      \vspace{-0.05in}
  \begin{equation}
    s_i^{+}\leq \frac{\frac{p_+'D(p'_+)}{S_{-i}^{+}+s_i^{-}}-p_-'}{\frac{D(p'_+)}{S_{-i}^{+}+s_i^{+}}-p'_-}s_i^{-}\leq s_i^{-}, \nonumber
      \vspace{-0.05in}
  \end{equation}
  which also gives a contradiction.
\end{itemize}

(c) Now we show Part (iii), i.e., $D(p)\leq S(p)$ for $p\geq p'$. Recall that when $p=p'$,
\begin{equation}
U_{i}(x_{i})=f_{i}(x_{i})+p's_{i}-cx_{i}-cs_{i}, \label{eq:prop_p0}
\end{equation}
and the best response is denoted by $(x_{i}^*(p'), s_{i}^*(p'))$.
Assume the contradiction that for some $p_1>p'$ we have $D(p_1)>S(p_1)$. Then,  owner $i$'s utility function at $p_1$ becomes:
\begin{equation}
U_{i}(x_{i})=f_{i}(x_{i})+p_1s_{i}-cx_{i}-cs_{i}, \label{eq:prop_p1}
\end{equation}
where the best response is $(x_{i}^*(p_1), s_{i}^*(p_1))$.
Compare equation (\ref{eq:prop_p0}) and (\ref{eq:prop_p1}),  we have:
\begin{align}
&f_i(x_{i}^*(p'))+p's_{i}^*(p')-cs_{i}^*(p')-cx_{i}^*(p')\nonumber\\
\geq & f_i(x_{i}^*(p_1))+p's_{i}^*(p_1)-cs_{i}^*(p_1)-cx_{i}^*(p_1) \label{eq:prop_supply1}
\end{align}
and
\begin{align}
&f_i(x_{i}^*(p_1))+p_1s_{i}^*(p_1)-cs_{i}^*(p_1)-cx_{i}^*(p_1)\nonumber\\
\geq & f_i(x_{i}^*(p'))+p_1s_{i}^*(p')-cs_{i}^*(p')-cx_{i}^*(p') \label{eq:prop_supply2}
\end{align}
Summing  (\ref{eq:prop_supply1}) and (\ref{eq:prop_supply2}), we have:
\begin{equation}
p_1(s_{i}^*(p_1)-s_{i}^*(p'))\geq p'(s_{i}^*(p_1)-s_{i}^*(p')).
\end{equation}
Since $p_1>p'>0$, we have:
\begin{equation}
s_{i}^*(p_1)-s_{i}^*(p')\geq 0.
\end{equation}
Since $i$ is arbitrary, we have $S(p_1)\geq S(p')$, and $S(p_1)\geq S(p')= D(p')\geq D(p_1)$ by Proposition \ref{prop:demand}, which leads to a contradiction.

(d) Now consider Part (i). From the above, if we  let $p_c$ be the smallest market clearing price, $p_c=\min\{p'\,|\,D(p')=S(p')\}$, we have $S(p)\geq D(p), \,\,\, \forall p\geq p_c$. 
The nondecreasing property of $S(p)$ for $p\leq p'$ can be shown as follows. Since $D(p)\leq S(p)$ for $p\leq p'$,  the utility function  for each owner $i$ is given by:
\begin{equation}
U_{i}(x_{i},s_{i})=f_{i}(x_{i})+ps_{i}-cx_{i}-cs_{i}. \nonumber
\end{equation}
Thus, each one's $s_i$ is nondecreasing when $p$ increases.

(e) Finally, Part (iv)  follows since at price $p_{upper}$, the market is not profitable, i.e., $p_{upper}\min\{\frac{D(p_{upper})}{S(p_{upper}}, 1\})\leq p_{upper} D(p_{upper})=c$. Thus, $s_i=0$ for all $i$.
\end{proof}

\noindent \textbf{Proof of Uniqueness}

We prove uniqueness of Nash equilibrium here, given that  $f(\cdot)$ is differentiable.

First, if $p<p_c$, then $S(p)<D(p)$. In this case each owner is doing optimization independently by solving following utility maximization problem:
\begin{align*}
 \max_{x_i, s_i}& \quad  U_{i}(x_{i},s_{i})=f_{i}(x_{i})+ps_{i}-cx_{i}-cs_{i} \\
 \text{s.t.}& \quad x_i\geq 0, \,\,  s_i\geq 0,\,\,  s_{i}+x_{i}\leq 1. \label{eq:optmize}
\end{align*}
Thus, the solution is unique for everyone due to the strict concavity of utility functions.

Now consider $p\geq p_c$. We prove the result by contradiction.
Suppose there exist more than one equilibrium state.  
Denote $\Omega(p)=\{\omega(p)=(\mathbf{x}^*(p),\mathbf{s}^*(p))\}$ the set of  Nash equilibria of the system given $p$.
We first introduce the following lemma before proving uniqueness:

\begin{lemma}\label{lemma:gradient}
Let $\Lambda=\{p|D(p)=S(p),p\in [p_{c},p_{upper}]\}$ denote the set of prices in $[p_{c},p_{upper}]$ such that demand equals supply in, we have following results.

(a) When $p \notin \Lambda$ and market reaches Nash equilibrium, we must have:
\begin{equation}
  \partial_{+} f_i(x_i^*(p))\leq \frac{\mathbf{d}Q_i(s_i;p)}{\mathbf{d}s_i}\bigg|_{s_i^*}\leq  \partial_{-} f_i(x_i^*(p)) \label{eq:lemma_gradient}
\end{equation}
for any owner $i$ such that $0<x_i^*(p), s_i^*(p)<1$,  where $Q_i(s_i;p)=p\frac{D(p)}{S(p)}s_i$.
If $s_i^*(p)=0$, the second inequality of (\ref{eq:lemma_gradient}) holds (the first holds if $x_i^*(p)=0$).
If $f_i(x_i)$ is differentiable, $\frac{\mathbf{d}f_i(x_i)}{\mathbf{d}x_i}\bigg|_{x_i^*}=\frac{\mathbf{d}Q_i(s_i;p)}{\mathbf{d}s_i}\bigg|_{s_i^*}$.

(b) When $p\in\Lambda$, the second inequality of (\ref{eq:lemma_gradient}) holds and we have:
\begin{equation}
  p\geq \partial_{+} f_i(x_i^*(p))\label{eq:lemma_gradient2}
\end{equation}
\end{lemma}

Lemma \ref{lemma:gradient} shows that when $f_i(x_i)$ is differentiable, the derivative of $f_i(x_i^*)$ equals derivative of $Q_i(s_i^*, p)$ when $s_i^*<1$, and the derivative of $f_i(x_i^*)$ is less than  derivative of $Q_i(s_i^*, p)$ when $s_i^*=1$.
There are  three situations:
\begin{enumerate}
  \item $\forall \omega_{k} \in \Omega$, we have $S^{\omega_{k}}(p)=D^{\omega_{k}}(p)$.
        This is impossible due to the similar reason as showed above in the case when $p<p_c$. In this case each owner is doing concave optimization independently and thus yields only one solution.
  \item $\exists \omega_{k} \in \Omega$ such that $S^{\omega_{k}}(p)=D^{\omega_{k}}(p)$, and $\exists \omega_{j} \in \Omega$ such that $S^{\omega_{j}}(p)>D^{\omega_{j}}(p)$. We have $S^{\omega_{j}}(p)>S^{\omega_{k}}(p)$ and there must exists some $i$ such that $s_i^{\omega_{j}}(p)>s_i^{\omega_{k}}(p)$. In this case, we have by Part (a) of Lemma \ref{lemma:gradient} that
       \begin{align*}
       p > & \frac{pD(p)S_{-i}^{\omega_{j}}}{(S^{\omega_{j}})^2}
       =\frac{\mathbf{d}Q_i(s_i;p)^{\omega_{j}}}{\mathbf{d}s_i}\bigg|_{s_i^{\omega_{j}}}\\
       = & \frac{\mathbf{d}f_i(x_i)}{\mathbf{d}x_i}\bigg|_{x_i^{\omega_{j}}}\\
       \geq & \frac{\mathbf{d}f_i(x_i)}{\mathbf{d}x_i}\bigg|_{x_i^{\omega_{k}}}\\
       =& p
       \end{align*}
Here the second inequality is due to the fact that $x_i^{\omega_{j}} \leq x_i^{\omega_{k}}$. For otherwise $x_i^{\omega_{j}} > x_i^{\omega_{k}}$ and $s_i^{\omega_{j}}(p)>s_i^{\omega_{k}}(p)$. Then, at equilibrium $\omega_{k}$, user $i$ should have used the larger $x_i^{\omega_{j}}$ (this is possible since $x_i^{\omega_{k}}+s_i^{\omega_{j}}<1$).
This yields contradiction.

  \item $\forall \omega_{k} \in \Omega$, we have $S^{\omega_{k}}(p)>D^{\omega_{k}}(p)$.
       In this case, all owners' utility functions  are given by:
       \begin{equation*}
         U_{i}(x_{i},s_{i})=f_{i}(x_{i})+\frac{D(p)}{S(p)}s_{i}-cx_{i}-cs_{i}.
       \end{equation*}
       Let $\mathbf{z}_i=[x_i, s_i]^{T} $denote action vector of $i$, and $\mathbf{z}=[\mathbf{z_1},...,\mathbf{z_{N_O}}]^T$.
       Note that $\mathbf{z}\in \mathbf{A}$, where the action space  $\mathbf{A}=\{(x,s)|x,s\geq 0, x+s\leq 1\}$, i.e., and
       $\mathbf{A}$ is a convex compact set.
      Denote
       \begin{equation*}
         h(\mathbf{z})=[\nabla_1 U_i(x_1,s_1), ..., \nabla_{N_O} U_i(x_{N_O},s_{N_O})]^{T},
       \end{equation*}
       where $\nabla_i U_i(x_i,s_i)=[\frac{\partial U_i(x_i,s_i)}{\partial x_i}  \frac{\partial U_i(x_i,s_i)}{\partial s_i}]^T$ is the gradient of $U_i(x_i,s_i)$ with respect to $\mathbf{z}_i$ at $\mathbf{z}$.
       Below we prove that $\sum_{i\in\mathcal{O}}U_i(\mathbf{z}_i)$ is diagonal strictly concave \cite{rosen1965} for any $\bar{\mathbf{z}}$, $\mathbf{z}^*\in\mathbf{A}$. Then, by  \cite{rosen1965}, the Nash equilibrium is unique. To this end, we have:
       \begin{align*}
         &(\bar{\mathbf{z}}-\mathbf{z}^*)^{T}(h(\mathbf{z}^*)-h(\bar{\mathbf{z}})) \\
         =&\sum_{i\in\mathcal{O}}(\bar{x_i}-x_i^*)(f'(x_i^*)-f'(\bar{x_i}))\\
         &+pD(p)\sum_{i\in\mathcal{O}}(\bar{s_i}-s_i^*)\bigg(\frac{S_{-i}^{*}(p)}{(S^{*}(p))^2}-\frac{\bar{S}_{-i}(p)}{(\bar{S}(p))^2}\bigg)\\
         >&pD(p)\sum_{i\in\mathcal{O}}(\bar{s_i}-s_i^*)\bigg(\frac{S_{-i}^{*}(p)}{(S^{*}(p))^2}-\frac{\bar{S}_{-i}(p)}{(\bar{S}(p))^2}\bigg)\\
         =&pD(p)\sum_{i\in\mathcal{O}}(\bar{s_i}-s_i^*)(\frac{1}{S^*}-\frac{1}{\bar{S}}-\frac{s_i^*}{(S^*)^2}+\frac{\bar{s}_i}{(\bar{S})^2})\\
         =&pD(p)\frac{(\bar{S}-S^*)^2}{S^{*}\bar{S}}+pD(p)\sum_{i\in\mathcal{O}}(\bar{s_i}-s_i^*)(\frac{\bar{s}_i}{(\bar{S})^2}-\frac{s_i^*}{(S^*)^2})\\
         =&pD(p)\frac{(\bar{S}-S^*)^2}{S^{*}\bar{S}}\\
         &+pD(p)\sum_{i\in\mathcal{O}}[(\frac{\bar{s}_i}{\bar{S}}-\frac{s_i^*}{S^*})^2
         +\frac{2\bar{s}_{i} s_{i}^{*}}{\bar{S}S^*}-\frac{\bar{s}_{i} s_{i}^{*}}{(\bar{S})^2}-\frac{\bar{s}_{i} s_{i}^{*}}{(S^*)^2}]\\
         =&pD(p)[\frac{(\bar{S}-S^*)^2}{S^{*}\bar{S}}+(\sum_{i\in\mathcal{O}}(\frac{\bar{s}_i}{\bar{S}}-\frac{s_i^*}{S^*})^2-\frac{\bar{s}_{i} s_{i}^{*}}{(\bar{S}S^*)^2}(\bar{S}-S^*)^2)]\\
         =& pD(p)[\sum_{i\in\mathcal{O}}(\frac{\bar{s}_i}{\bar{S}}-\frac{s_i^*}{S^*})^2+\frac{(\bar{S}-S^*)^2}{S^{*}\bar{S}}(1-\sum_{i\in\mathcal{O}}\frac{\bar{s}_{i} s_{i}^{*}}{\bar{S}S^*})]\\
         \geq & pD(p)[\sum_{i\in\mathcal{O}}(\frac{\bar{s}_i}{\bar{S}}-\frac{s_i^*}{S^*})^2+\frac{(\bar{S}-S^*)^2}{S^{*}\bar{S}}(1-\sum_{i\in\mathcal{O}}\frac{\bar{s}_{i} }{\bar{S}})]\\
         >& 0.
       \end{align*}
The second to last inequality is due to the fact that $s_i^*/S^*\leq1$.
We have then by  \cite{rosen1965} that the equilibrium is unique.
\end{enumerate}
From the three cases above, we conclude  the uniqueness of Nash equilibrium.

\subsection{Proof of Lemma \ref{lemma:gradient}}
For simplicity of representation, we ignore parameter $p$ in function $Q_i(s_i;p)$ since $p$ is fixed.

We first prove Part (a), i.e., consider  $p\notin \Lambda$. Note that in this case $D(p)<S(p)$.

(i) We look at the left inequality of equation (\ref{eq:lemma_gradient}).
Suppose instead $\frac{\mathbf{d}Q_i(s_i^*)}{\mathbf{d}s_i}<\partial_{+} f_i(x_i^*)$.
For $0<s_i^*(p)\leq 1$, we have:
\begin{align}
  &\lim_{\delta\rightarrow 0^+} U_i(x_i^*+\delta,s_i^*-\delta)-U_i(x_i^*,s_i^*) \nonumber\\
  =\,& \lim_{\delta\rightarrow 0^+} f_i(x_i^*+\delta)-f_i(x_i^*)+ Q_i(s_i^*-\delta)-Q_i(s_i^*) \nonumber\\
  =\,&\lim_{\delta\rightarrow 0^+} (\partial_{+} f_i(x_i^*) -\frac{\mathbf{d}Q_i(s_i^*)}{\mathbf{d}s_i})\delta \nonumber\\
  \geq\, & 0. \nonumber
\end{align}
Thus, decreasing supply by an infinitesimal $\delta$ leads to a higher utility, which gives a contradiction. Note that analysis in this step allows $x_i^*=0$.

(ii) Now we turn to the right inequality of (\ref{eq:lemma_gradient}). Similarly suppose $\frac{\mathbf{d}Q_i(s_i^*)}{\mathbf{d}s_i}> \partial_{-} f_i(x_i^*)$ for $p\geq p_c$ and $0<x_i^*(p)\leq 1$. We have a similar proof:
\begin{align}
  &\lim_{\delta\rightarrow 0^+} U_i(x_i^*-\delta,s_i^*+\delta)-U_i(x_i^*,s_i^*) \nonumber\\
  =\,& \lim_{\delta\rightarrow 0^+} f_i(x_i^*-\delta)-f_i(x_i^*)+ Q_i(s_i^*+\delta)-Q_i(s_i^*) \nonumber\\
  =\,&\lim_{\delta\rightarrow 0^+} (- \partial_{-} f_i(x_i^*) +\frac{\mathbf{d}Q_i(s_i^*)}{\mathbf{d}s_i})\delta \nonumber\\
  \geq\, & 0. \nonumber
\end{align}
This also gives a contradiction. Note that analysis in this step allows $s_i^*=0$.

(iii) When $f_i(x_i)$ is differentiable, the analysis still applies.

(b) Now we consider $p\in \Lambda$. Note that $D(p)=S(p)$ in this case.
In this case, the proof of second inequality of (\ref{eq:lemma_gradient})
 is the same as  case (ii) above. This is so because although $D(p)=S(p)$, when consider increasing $s_i^*$ by a small $\delta$.

As for  inequality (\ref{eq:lemma_gradient2}), suppose $\frac{\mathbf{d}Q_i(s_i^*)}{\mathbf{d}s_i}<\partial_{+} f_i(x_i^*)$. Note that $Q_i(s_i^*;p)=p_c\frac{D(p)}{S(p)}=p$. For $0<s_i^*(p)\leq 1$, we have:
\begin{align}
  &\lim_{\delta\rightarrow 0^+} U_i(x_i^*+\delta,s_i^*-\delta)-U_i(x_i^*,s_i^*) \nonumber\\
  =\,& \lim_{\delta\rightarrow 0^+} f_i(x_i^*+\delta)-f_i(x_i^*)+ p(s_i^*-\delta)-Q_i(s_i^*) \nonumber\\
  =\,&\lim_{\delta\rightarrow 0^+} (\partial_{+} f_i(x_i^*) -p)\delta \nonumber\\
  \geq\, & 0, \nonumber
\end{align}
which gives a contradiction.

Combining (a) and (b) we have   the lemma.

\subsection{Proof of Proposition \ref{prop:demand}}
\begin{proof}
Under the concavity and continuity assumption of $g_k$,   a renter has continuous optimal demand $y_k^*(p)\in[0,1]$ which maximizes his overall utility $U_k(y_k)$ given by (\ref{eq:utilty_renter}).

 Suppose the market price is  $p_{0}$, we must have:
\begin{align*}
   &U_k^{p_0}(y_k^*(p_0))- U_k^{p_0}(y_k') \\
   =& g_k(y_k^*(p_0))-p_{0}y_k^*(p_0)-g_k(y_k')+p_{0}y_k'\\
   \geq & 0, \quad \forall k_i' \in[0,1]
\end{align*}
 That is,
 \begin{equation*}
   g_k(y_k^*(p_0))-g_k(y_k'))\geq p_{0}(y_k^*(p_0)-y_k'), \quad \forall y_k'\in [0,1]
 \end{equation*}
Hence, when price is $p_1$ such that $p_1>p_0$, for any $y_k'>y_k^*(p_0)$, we must have:
\begin{align*}
   &U_k^{p_1}(y_k^*(p_0))- U^{p_1}_i(y_k') \\
   =& g_k(y_k^*(p_0))-p_{1}y_k^*(p_0)-g_k(y_k')+p_{1}y_k'\\
   \geq & (p_{0}-p_{1})(y_k^*(p_0)-y_k')\\
   >& 0
\end{align*}
Here the first inequality follows from the above equation.
 This implies $y_k^*(p_1)\leq y_k^*(p_0)$, i.e., each renter's demand is non-increasing in $p$.

For any renter $k$, we have $g_k(y_k)<By_{k}, y_k\in[0,1]$ since  derivative of $g_k(y_k)$ is bounded by $B$. Then, every renter has zero demand when $p\geq B$.
\end{proof}

\subsection{Proof of Theorem \ref{thm:revenue-supply}}
Denote owner $i$'s strategy at the market equilibrium under price $p$ as $x_i^*(p), s_i^*(p)$.
Before proving Theorem \ref{thm:revenue-supply}, we introduce the following lemmas (here $\partial_{+} f_i(x_i^*(p)),\partial_{-} f_i(x_i^*(p))$ denote the right and left derivatives of $f_i(x_i)$ at $x_i^*$, which always exist since $f_i(x_i)$ is concave).
\begin{lemma}\label{lemma:2owner}
When $p\in[p_c,p_{upper}]$ and $S(p)>0$, there exist at least two $i\in \mathcal{O}$ such that $s_i>0$.
\end{lemma}
\begin{lemma}\label{lemma:full}
When $c<p\leq p_c$, we have $x_i^*(p)+s_i^*(p)=1$ for all $i\in \mathcal{O}$.
\end{lemma}
\begin{lemma}\label{lemma:full_after}
Given price $p>p_c$, for any $i$ such that $s_i^*(p)> s_i^*(p_c)$, we have $x_i^*(p)+s_i^*(p)=1$ and $x_i^*(p)<x_i^*(p_c)$.
\end{lemma}

We prove the theorem by contradiction.
Let $p_1,p_2>p_c$ be such that $p_1D(p_1)\geq p_2(D(p_2)$ and $D(p_1)<S(p_1)$ and $D(p_2)<S(p_2)$.
Suppose that $S(p_1)<S(p_{2})$.

Let $\mathcal{K}$ denote the set of owners such that $s_i^*(p_{1})<s_i^*(p_{2})\leq 1$ for all $i\in \mathcal{K}$, i.e., $\mathcal{K}=\{i|s_i^*(p_{1})<s_i^*(p_{2})\}$.
Since $S(p_1)<S(p_{2})$, $\mathcal{K}\neq\emptyset$. Since $f_i(x_i)$ is concave, we know that for $i\in\mathcal{K}$, $x_i^*(p_{1})\geq x_i^*(p_{2})$ (otherwise owner $i$ can increase $x_i^*(p_{1})$ to $x_i^*(p_{2})$).

(a) We first consider the case when $x_i^*(p_{1})> x_i^*(p_{2}), \,\forall i\in \mathcal{K}$.
We consider the following cases.
\begin{enumerate}
  \item If there exists some $i\in \mathcal{K}$ such that $S_{-i}(p_{1})\geq S_{-i}(p_{2})$. Then,  we have for $i$ that:
    \begin{align*}
      &\frac{\mathbf{d}Q_i(s_i^*(p_{1});p_{1})}{\mathbf{d}s_i}=\frac{p_{1}D(p_{1})S_{-i}(p_{1})}{(S_{-i}(p_{1})+s_i^*(p_{1}))^2} \\
      > &\frac{p_{2}D(p_{2})S_{-i}(p_{2})}{(S_{-i}(p_{2})+s_i^*(p_{2}))^2}\geq \partial_{+}f_i(x_i^*(p_2)) \\
      \geq & \partial_{-}f_i(x_i^*(p_1)).
    \end{align*}
    Here the first inequality is obtained by the condition in this step that  $S_{-i}(p_{1})\geq S_{-i}(p_{2})$ and the assumption that $S_{-i}(p_{1})+s_i^*(p_{1})=S(p_1)<S(p_{2})=S_{-i}(p_{2})+s_i^*(p_{2})$. From Lemma \ref{lemma:gradient} we know that $x_i^*(p_1), s_i^*(p_1)$ will not be an equilibrium strategy, contradiction.
  \item If no $i$ satisfies the condition in step 1, we have $S_{-i}(p_{1})< S_{-i}(p_{2}),\, \forall i \in\mathcal{K}$. We now consider if there exists $i\in \mathcal{K}$ satisfies $S_{-i}(p_1)\geq s_{i}^*(p_{1})$, i.e., $s_i^*$ is smaller than others' sharing at $p_{1}$. For such an owner $i$, we have:
      \begin{align*}
        &\frac{\mathbf{d}Q_i(s_i^*(p_{1});p_{1})}{\mathbf{d}s_i}=\frac{p_{1}D(p_{1})S_{-i}(p_{1})}{(S_{-i}(p_{1})+s_i^*(p_{1}))^2} \\
        > & \frac{p_{2}D(p_{2})S_{-i}(p_{2})}{(S_{-i}(p_{2})+s_i^*(p_{1}))^2}>\frac{p_{2}D(p_{2})S_{-i}(p_{2})}{(S_{-i}(p_{2})+s_i^*(p_{2}))^2} \\
        \geq &  \partial_{+}f_i(x_i^*(p_2))\geq \partial_{-}f_i(x_i^*(p_1)).
      \end{align*}
      Here the first inequality is given by the fact that function $q(x)=\frac{x}{(x+k)^2}$ is monotone decreasing in $x\geq k$ and $p_1D(p_1)\geq p_2D(p_2)$, and the second inequality is from the definition of $\mathcal{K}$.

  \item If no $i$ satisfied any conditions in above two steps, i.e., $S_{-i}(p_{1})< S_{-i}(p_{2})$ and $S_{-i}(p_1)< s_{i}^*(p_{1})$ for all $i \in\mathcal{K}$.
      Note that $S_{-i}(p_1)< s_{i}^*(p_{1})$ implies there exists at most one $i$ in $\mathcal{K}$ in this case. Since $\mathcal{K}\neq\emptyset$, there is exactly one $i'\in \mathcal{K}$. By the definition of $\mathcal{K}$, we have $s_j^*(p_1)>s_j^*(p_{2}), \, \forall j\neq i$.
      However, we have $S_{-i}(p_{1})< S_{-i}(p_{2})$, which gives a contradiction.
\end{enumerate}

(b) Now we consider the case when there exists $i\in\mathcal{K}$ such that $x_i^*(p_{1})= x_i^*(p_{2})$.  For such an $i$, we have:
\begin{equation}
  \partial_{+}f_i(x_i^*(p_{1}))\leq c = \frac{\mathbf{d}Q_i(s_i^*(p_{1});p_{1})}{\mathbf{d}s_i}\leq  \partial_{-}f_i(x_i^*(p_{1})). \label{eq:foo-1}
\end{equation}
The first inequality is due to the fact that $x_i^*(p_{1}) + s_i^*(p_{1})<x_i^*(p_{1}) + s_i^*(p_{2})\leq1$ so that the partial derivative of $f_i$ is no more than the cost, whereas the second inequality is due to Lemma \ref{lemma:gradient} and that $s_i^*(p_{1})<s_i^*(p_{2})\leq1$.

We can adopt the same argument as above to obtain:
\begin{align*}
  &\frac{\mathbf{d}Q_i(s_i^*(p_{1});p_{1})}{\mathbf{d}s_i} \\
      = & \frac{p_{r}D(p_{1})S_{-i}(p_{1})}{(S_{-i}(p_{1})+s_i^*(p_{1}))^2}> \frac{p_{2}D(p_{2})S_{-i}(p_{2})}{(S_{-i}(p_{2})+s_i^*(p_{2}))^2}.
\end{align*}
Note that the last term is exactly $\frac{\mathbf{d}Q_i(s_i^*(p_{2});p_{2})}{\mathbf{d}s_i}$, which satisfies $\frac{\mathbf{d}Q_i(s_i^*(p_{2});p_{2})}{\mathbf{d}s_i}\geq c$.
Combining this with the above, we have:
\begin{align*}
\frac{\mathbf{d}Q_i(s_i^*(p_{1});p_{1})}{\mathbf{d}s_i} > c,
\end{align*}
which contradicts with (\ref{eq:foo-1}).

Combining (a) and (b)  we have $S(p_{1})\geq S(p_{2})$.

\subsection {Proof of Lemma\ref{lemma:2owner}}
Suppose there exists only one owner $i$ such that $s_i>0$.
Since $p\geq p_c$, we have $S(p)\geq D(p)$ by Proposition \ref{prop:demand}. Meanwhile, we must have $s_i\leq D(p)$ given utility in equation (\ref{eq:utility_owner}) since $S(p)=s_i$. Thus, we have $D(p)=S(p)=s_i$. Then consider two cases,

(i) If $p<p_o$, we have $D(p)>1$, since $D(p)=s_i\leq 1$, contradiction.

(ii) If $p\geq p_o$, for any $j$ such that $s_j=0$, according to Lemma \ref{lemma:gradient}, we have:
$p_o>\partial_{-} f_j(1) \geq \frac{\mathbf{d}Q_j(s_j;p)}{\mathbf{d}s_j}\bigg|_{s_j^*=0}= p_o$, where the first inequality is due to Assumption \ref{assumption:utility-function}. This gives a contradiction.

\subsection{Proof of Lemma \ref{lemma:full}}
(a) If $c<p<p_c$, we have $x_i^*(p)+s_i^*(p)=1$ for all $i\in\mathcal{O}$. Otherwise suppose for some $i$ such that  $x_i^*(p)+s_i^*(p)<1$. Then, we will have
\begin{align*}
&\lim_{\delta\rightarrow 0^+}U_i(x_i^*(p),s_i^*(p)+\delta)-U_i(x_i^*(p),s_i^*(p))\\
=& \lim_{\delta\rightarrow 0^+}(p-c)\delta \\
\geq & 0,
\end{align*}
which is a contradiction.

(b)If $p=p_c$. Let $x_i^\delta,s_i^\delta$ denote $i$'s strategy at equilibrium at $p_\delta=p_c-\delta>c$.
From above analysis, we have $x_i^\delta+s_i^\delta=1$.
Consider some strategy $(x_i^\Delta,s_i^\Delta)$ such that $\Delta=1-x_i^\Delta-s_i^\Delta>0$.
We have:
\begin{equation*}
  f_i(x_i^\delta)+p_\delta s_i^\delta -cs_i^\delta-cx_i^\delta \geq f_i(x_i^\Delta)+p_\delta s_i^\Delta -cs_i^\Delta-cx_i^\Delta
\end{equation*}
that gives
\begin{equation*}
  f_i(x_i^\delta)+p_{c} s_i^\delta -\delta s_i^\delta \geq f_i(x_i^\Delta)+p_{c} s_i^\Delta -\delta s_i^\Delta+c\Delta.
\end{equation*}
Hence,
\begin{align*}
  & \lim_{\delta\rightarrow 0^+}U_i^{p=p_c}(x_i^\delta,s_i^\delta)-U_i^{p=p_c}(x_i^\Delta,s_i^\Delta)\\
  \geq & \lim_{\delta\rightarrow 0^+}-\delta +c\Delta\\
  > & 0.
\end{align*}
That is, strategy $(x_i^\delta,s_i^\delta)$ outperforms all strategies $(x_i,s_i)$ such that $x_i+s_i<1$, which means $x_i^*(p_c)+s_i^*(p_c)=1$.
\subsection{Proof of Lemma \ref{lemma:full_after}}
Given $p>p_c$, consider $i$ such that $s_i^*(p)>s_i^*(p_c)\geq 0$. From Lemma \ref{lemma:full} we have $x_i^*(p_c)+s_i^*(p_c)=1$.
Since $s_i^*(p)>s_i^*(p_c)$, $x_i^*(p)\leq 1-s_i^*(p)<1-s_i^*(p_c)=x_i^*(p_c)$, i.e, $x_i^*(p)<x_i^*(p_c)$.

Note that $c\leq\partial_{-}f_i(x_i^*(p_c))\leq \partial_{+}x_i^*(p)$, $f_i(x_i)-cx_i$ is increasing in $[0,x_i^*(p_c)]$. So we must have $f_i(1-s_i^*(p))-c(1-s_i^*(p))\geq f_i(x_i)-cx_i$, for any $x_i\in[0,(1-s_i^*(p)]$.
So far we have proved the lemma.

\subsection{Proof of Theorem \ref{thm:pr_psw}}
We first give the result that both $p_{sw}$ and $p_{r}$ are inside region $[p_{c},p_{upper}]$.

\begin{prop}\label{prop:sw}
The following results hold.
\begin{itemize}
\item The social welfare under any $p<p_c$ is no more than that under $p_c$. 
\item Under revenue maximization policies,  supply is always no less than demand,  i.e.,  $S(p_{r})\geq D(p_{r})$.
\end{itemize}
\end{prop}

(a) We prove the first bullet in Theorem \ref{thm:pr_psw} here. For any $p_1>p_c$, we have $S(p_c)=D(p_c)$, $S(p_1)\geq D(p_1)$.
We want to show that:
\begin{align*}
    &\sum_{i\in \mathcal{O}}U_i(x_i^*(p_1),s_i^*(p_1))+\sum_{k\in \mathcal{R}}U(y_k^*(p_1)) \\
    -&\sum_{i\in \mathcal{O}}U_i(x_i^*(p_c),s_i^*(p_c))+\sum_{k\in \mathcal{R}}U(y_k^*(p_c))\leq 0
\end{align*}
Note that for $p\geq p_c$ (the revenue terms cancel out):
\begin{align}
  &\sum_{i\in \mathcal{O}}U_i(x_i^*(p),s_i^*(p))+\sum_{k\in \mathcal{R}}U(y_k^*(p))\nonumber\\
  =&\sum_{i\in \mathcal{O}}[f_i(x_i^*(p))-cs_i-cx_i]+\sum_{k\in \mathcal{R}}g_k(y_k^*(p)).\label{eq:welfare-foo}
\end{align}
For any renter $k$, we have:
\begin{equation*}
   g_k(y_k^*(p_1))-p_{c}y_k^*(p_1)\leq g_k(y_k^*(p_c))-p_{c}y_k^*(p_c)
\end{equation*}
hence,
\begin{equation*}
  g_k(y_k^*(p_1))-g_k(y_k^*(p_c))\leq p_{c}y_k^*(p_1)-p_{c}y_k^*(p_c)
\end{equation*}
sum over all renters we have:
\begin{equation}\label{eq:renter_sw_lost}
  \sum_{k\in \mathcal{R}}g_k(y_k^*(p_1))-g_k(y_k^*(p_c))\leq p_{c}D(p_1)-p_{c}D(p_c)
\end{equation}

For owners, we discuss in two cases:
\begin{enumerate}
  \item For $i$ such that $s_i^*(p_1) \leq s_i^*(p_c)$, we have:
  \begin{align*}
    &f_i(x_i^*(p_1))+p_{c}s_i^*(p_1)-cs_i^*(p_1)-cx_i^*(p_1)\\
    \leq  & f_i(x_i^*(p_c))+p_{c}s_i^*(p_c)-cs_i^*(p_c)-cx_i^*(p_c)
  \end{align*}
  thus,
  \begin{align*}
    &f_i(x_i^*(p_1))-f_i(x_i^*(p_c)) \\
    &-cx_i^*(p_1)-cs_i^*(p_1)+cx_i^*(p_c)+cs_i^*(p_c) \\
    \leq & p_c(s_i^*(p_c)-s_i^*(p_1))
  \end{align*}
  \item For $i$ such that $s_i^*(p_1) > s_i^*(p_c)$, from Lemma \ref{lemma:full_after} we must have $x_i^*(p_1)+s_i^*(p_1)=1$ and $0\leq x_i^*(p_1)<x_i^*(p_c)$.
Thus,
  \begin{align*}
    &f_i(x_i^*(p_1))-f_i(x_i^*(p_c)) \\
    &-cx_i^*(p_1)-cs_i^*(p_1)+cx_i^*(p_c)+cs_i^*(p_c) \\
    = & f_i(x_i^*(p_1))-f_i(x_i^*(p_c))\\
    \leq & \partial_{-}f_i(x_i^*(p_c))(x_i^*(p_1)-x_i^*(p_c))\\
    \leq & p_c(x_i^*(p_1)-x_i^*(p_c))\\
    = & p_c[(1-s_i^*(p_1))-(1-s_i^*(p_c)))]\\
    = & p_c(s_i^*(p_c)-s_i^*(p_1))
  \end{align*}
  Here the inequality is due to the concavity of $f$.
Summing above two cases, we have:
\begin{align}\label{eq:owner_sw_lost}
  &\sum_{i\in\mathcal{O}}[f_i(x_i^*(p_1))-f_i(x_i^*(p_c))\nonumber\\
  &-cx_i^*(p_1)-cs_i^*(p_1)+cx_i^*(p_c)+cs_i^*(p_c)]  \nonumber\\
  \leq & \sum_{s_i^*(p_1) \leq s_i^*(p_c)}p_c(s_i^*(p_c)-s_i^*(p_1))\nonumber\\
  &+\sum_{s_i^*(p_1) > s_i^*(p_c)}p_c(s_i^*(p_c)-s_i^*(p_1))\nonumber\\
  =&p_c(S(p_c)-S(p_1))
\end{align}
\end{enumerate}
Summing up (\ref{eq:renter_sw_lost}) and (\ref{eq:owner_sw_lost}) and using (\ref{eq:welfare-foo}), we have:
\begin{align*}
    &\sum_{i\in \mathcal{O}}U_i(x_i^*(p_1),s_i^*(p_1))+\sum_{k\in \mathcal{R}}U(y_k^*(p_1)) \\
    -&\sum_{i\in \mathcal{O}}U_i(x_i^*(p_c),s_i^*(p_c))+\sum_{k\in \mathcal{R}}U(y_k^*(p_c))\\
    \leq & p_{c}S(p_c)-p_{c}S(p_1)+p_{c}D(p_1)-p_{c}D(p_c)\\
    \leq & 0
\end{align*}

(b) By Proposition \ref{prop:sw} and results above, we must have $p_{r}\geq p_{sw}$. Part (ii) and Part (iii) also follow.


\subsection{Proof of Proposition \ref{prop:sw}}
\begin{proof}
(i) Consider the case when the price  is $p_0$ and $D(p_0)>S(p_0)$ (meaning $p_0<p_c$). There exists some renters who are not able to rent products since available products are not enough. Let $\mathcal{R}_0$ denote the set of renters who have successfully rented a product and $\mathcal{R}/\mathcal{R}_0$ denotes the other renters.

We know that $U_i(y_i)\geq0,\forall i\in \mathcal{R}_0$ and $U_i(y_i)=0, y_i=0,\forall i\notin \mathcal{R}_0, i\in R$.
According to Theorem \ref{thm:ne_exist}, there exists some market clearing price $p_c>p_0$ such that $D(p_c)=S(p_c)$.

For owners,  we must have each owner's  social welfare  $U_{i}(x_{i})=f_{i}(x_{i})+p_{c}s_{i}-cx_{i}-cs_{i}$ increases compared to the case when price is $p_0$.

For renters in $\mathcal{R}/\mathcal{R}_0$, each renter gains non-negative social welfare increment. For renters in $\mathcal{R}_0$, each renter $i$ suffers utility lost not exceeding $(p_{c}-p_0)y_i^*(p_0)$ even if they adopt the same usage  $y_i^*(p_0)$ as at $p_{c}$. Therefore total social welfare lost for group $\mathcal{R}_0$ will be bounded by $(p_{c}-p_0)S(p_0)$. However for the owners who have served users in $\mathcal{R}_0$, they have at least $(p_{c}-p_0)S(p_0)$ utility increment at price $p_c$ due to the price increase even if they do not increase supply.

To sum up, the social welfare at lowest market clearing price $p_{c}$, is no less than the social welfare at any other prices.
(ii) According to Theorem \ref{thm:ne_exist}, denote $p_{c}$ as the lowest market clearing price.
Consider any price $p_0<p_c$ and we have $D(p_0)>S(p_0)$. The volume of trade will be $p_0 S(p_0)$. According to Proposition \ref{prop:demand}, $D(p)$ is continuous. Thus, there exists some $p_1>p_{c}$ such that $S(p_0)=D(p_1)$ and the volume of trade will be $p_1 D(p_1)$ and we have $p_1D(p_1)>p_0S(p_0)$.
\end{proof}

\subsection{Proof of Theorem \ref{thm:gap}}
\begin{proof}
The first inequality is trivial. So we focus on the second inequality. Consider $p_{sw} =p_c$.

Denote $U_i^*(p)$ as $i$'s  optimal utility and $x_i^*(p)$, $s_i^*(p)$ (for $i\in
\mathcal{O}$) as his best response at price $p$.
Then, $x_i^*(p_{r})$, $s_i^*(p_{r})$ and $x_i^*(p_{sw})$, $s_i^*(p_{sw})$ are $i$'s best response usages and share levels (for $i\in\mathcal{O}$) at price $p_{r}$ and $p_{sw}$.
The total demand and total supply at price $p_{r}$ and $p_{sw}$ are given by:
\begin{equation*}
\begin{aligned}
D(p_{r})=\sum_{k\in\mathcal{R}}y_k^*(p_{r}),\hspace{0.1in} & S(p_{r})=\sum_{i\in\mathcal{O}}s_i^*(p_{r})\\
D(p_{sw})=\sum_{k\in\mathcal{R}}y_k^*(p_{sw}),\hspace{0.1in} & S(p_{sw})=\sum_{i\in\mathcal{O}}s_i^*(p_{sw})
\end{aligned}
\end{equation*}
Note that total social welfare is given by:
\begin{align*}
  &\sum_{i\in \mathcal{O}}U_i(x_i^*(p),s_i^*(p))+\sum_{k\in \mathcal{R}}U_k(y_k^*(p))\\
  =&\sum_{i\in \mathcal{O}}[f_i(x_i^*(p))-cs_i^*(p)-cx_i^*(p)]+\sum_{k\in \mathcal{R}}g_k(y_k^*(p)).
\end{align*}
We first consider renters' social welfare. Since $p_{r}\geq p_{sw}$,  for any renter $k$, we have $0 \leq y_k^*(p_{r})\leq y_k^*(p_{sw})$, and
\begin{align*}
  & g_k(y_k^*(p_{sw}))-g_k(y_k^*(p_{r}))\\
  \leq & \partial_{+} g_k(y_k^*(p_{r}))(y_k^*(p_{sw})-y_k^*(p_{r}))  \\
  \leq &  p_{r}(y_k^*(p_{sw})-y_k^*(p_{r}).
\end{align*}
The second inequality is obtained by the following. First, if $y_k^*(p_{r})<1$, then
\begin{equation*}
  \partial_{+}g_k(y_k^*(p_{r}))\leq p_{r},\, \forall k \in \mathcal{R}
\end{equation*}
This holds because $y_k^*(p_{r})$ is chosen to maximize $g_k(y_k)-p_ry_k$.
On the other hand, if $y_k^*(p_{r})=1$, then $y_k^*(p_{sw})=1$. Thus, the bound also holds.

Now summing up all renters' social welfare:
\begin{eqnarray}
\sum_{k\in\mathcal{R}} g_k(y_k^*(p_{sw}))-U_k(y_k^*(p_{r}))\leq &p_{r}(D(p_{sw})-D(p_{r})) \label{eq:renter_gap}
\end{eqnarray}
After bounding renters' social welfare, we now take a look at the owners. Note that by Lemma \ref{lemma:full}, we have $x_i^*(p_c)+s_i^*(p_c)=1$.
\begin{enumerate}
\item For owners such that $x_i^*(p_{r})\geq x_i^*(p_{sw})$, we have:
    \begin{align*}
        & f_i(x_i^*(p_{sw}))-cx_i^*(p_{sw})-cs_i^*(p_{sw})\\
        &-f_i(x_i^*(p_{r}))+cx_i^*(p_{r})+cs_i^*(p_{r}) \\
        \leq & f_i(x_i^*(p_{sw}))- f_i(x_i^*(p_{r}))\\
        \leq & \partial_{-}f_i(x_i^*(p_{r}))(x_i^*(p_{sw})-x_i^*(p_{r}))\\
        \leq & 0.
    \end{align*}
The last inequality holds since $\partial_{-}f_i(x_i^*(p_{r}))\geq0$.

\item For owners such that $x_i^*(p_{r})< x_i^*(p_{sw})$, we must have $s_i^*(p_{r})> s_i^*(p_{sw})$ and  $x_i^*(p_{r})+s_i^*(p_{r})=1$ from Lemma \ref{lemma:full_after}.
    Hence,
    \begin{align*}
        & f_i(x_i^*(p_{sw}))-cx_i^*(p_{sw})-cs_i^*(p_{sw})\\
        & -f_i(x_i^*(p_{r}))+cx_i^*(p_{r})+cs_i^*(p_{r}) \\
        = & f_i(x_i^*(p_{sw}))- f_i(x_i^*(p_{r}))\\
        \leq & \partial_{+}f_i(x_i^*(p_{r}))(x_i^*(p_{sw})-x_i^*(p_{r}))\\
        \leq & \frac{p_{r}D(p_{r})S_{-i}(p_{r})}{(S_{-i}(p_{r})+s_i^*(p_{r}))^2}(x_i^*(p_{sw})-x_i^*(p_{r}))\\
        \leq & \frac{p_{r}D(p_{r})}{S(p_{r})}(x_i^*(p_{sw})-x_i^*(p_{r}))\\
        = & \frac{p_{r}D(p_{r})}{S(p_{r})}(s_i^*(p_{r})-s_i^*(p_{sw}))
    \end{align*}
    The inequality here is because the marginal benefit from sharing should be at least as good as that from self-usage.
\end{enumerate}
Summing up above two cases over all owners , we have:
\begin{align}\label{eq:owner_gap_final}
&\sum_{i\in\mathcal{O}}f_i(x_i^*(p_{sw}))-cx_i^*(p_{sw})-cs_i^*(p_{sw})\nonumber\\
        & -f_i(x_i^*(p_{r}))+cx_i^*(p_{r})+cs_i^*(p_{r}) \nonumber\\
\leq & \frac{p_{r}D(p_{r})}{S(p_{r})}\sum_{s_i^r> s_i^{sw}}(s_i^*(p_{r})-s_i^*(p_{sw}))
\end{align}
Adding (\ref{eq:owner_gap_final}) and (\ref{eq:renter_gap})  together we have proved the theorem.
\end{proof}

\subsection{Proof of Theorem \ref{thm:subsidy-0}}
Suppose given $p$, let $x_i^{\epsilon_1}, s_i^{\epsilon_1}$ and $x_i^{\epsilon_2}, s_i^{\epsilon_2}$ denote $i$'s self usage and share at equilibrium, under $\epsilon_1$, and $\epsilon_2$, respectively.  Note that $0\leq \epsilon_1<\epsilon_2$.

We prove $S^{\epsilon_2}(p)\geq S^{\epsilon_1}(p)$ by contradiction with a similar argument used in proving Theorem \ref{thm:pr_psw}.
Similar to Lemma \ref{lemma:gradient}, we have:
\begin{equation}\label{eq:epsilon_lemma}
   \partial_{+} f_i(x_i^\epsilon)\leq \frac{\mathbf{d}Q_i(s_i;p)}{\mathbf{d}s_i}\bigg|_{s_i^\epsilon}+p\epsilon\leq  \partial_{-} f_i(x_i^\epsilon)
\end{equation}
for $i$ such that  $s_i^{\epsilon}<1$.
Suppose $S^{\epsilon_2}(p)< S^{\epsilon_1}(p)$ (note that in this case $\frac{pD(p)}{S^{\epsilon_1}}>\frac{pD(p)}{S^{\epsilon_2}}$). There must exists some $i$ such that his equilibrium supply $s_i^{\epsilon_2}<s_i^{\epsilon_1}$.

Let $\mathcal{K}=\{i|s_i^{\epsilon_2}<s_i^{\epsilon_1})\}$, we have $\mathcal{K}\neq\emptyset$.  Thus we must have $x_i^{\epsilon_2}>x_i^{\epsilon_1}, \forall i\in\mathcal{K}$. Otherwise we will have $U_i^{\epsilon_2}(x_i^{\epsilon_1},s_i^{\epsilon_1})\geq U_i^{\epsilon_2}(x_i^{\epsilon_2},s_i^{\epsilon_2})$.
We consider the following cases.
\begin{enumerate}
  \item If there exists some $i\in \mathcal{K}$ such that $S_{-i}^{\epsilon_2}\geq S_{-i}^{\epsilon_1}$, we have for $i$ that:
    \begin{align*}
      &\frac{\mathbf{d}Q_i(s_i^{\epsilon_2})}{\mathbf{d}s_i}+p\epsilon_2=\frac{pD(p)S_{-i}^{\epsilon_2}}{(S_{-i}^{\epsilon_2}+s_i^{\epsilon_2})^2}+p\epsilon_2 \\
      > &\frac{pD(p)S_{-i}^{\epsilon_1}}{(S_{-i}^{\epsilon_1}+s_i^{\epsilon_1})^2}+p\epsilon_1\geq \partial_{+}f_i(x_i^{\epsilon_1}) \\
      \geq & \partial_{-}f_i(x_i^{\epsilon_2}).
    \end{align*}
    Here the first inequality is obtained by   $S_{-i}^{\epsilon_2}\geq S_{-i}^{\epsilon_1}$ and $S_{-i}^{\epsilon_2}+s_i^{\epsilon_2}=S^{\epsilon_2}<S^{\epsilon_1}=S_{-i}^{\epsilon_1}+s_i^{\epsilon_1}$. From (\ref{eq:epsilon_lemma}) we know that $x_i^*(p_{r}), s_i^*(p_{r})$ will not be a equilibrium strategy. Contradiction.
  \item If no $i$ satisfies condition in step 1, we have $S_{-i}^{\epsilon_2}< S_{-i}^{\epsilon_1},\, \forall i \in\mathcal{K}$. We will need to check if there exists $i\in \mathcal{K}$ satisfies $S_{-i}^{\epsilon_2}\geq s_{i}^{\epsilon_2}$, i.e., $s_i^{\epsilon_2}$ is smaller than others' sharing at $p$ under subsidy. For such $i$, we have:
      \begin{align*}
        &\frac{\mathbf{d}Q_i(s_i^{\epsilon_2})}{\mathbf{d}s_i}+p\epsilon_2=\frac{pD(p)S_{-i}^{\epsilon_2}}{(S_{-i}^{\epsilon_2}+s_i^{\epsilon_2})^2}+p\epsilon_2 \\
        \geq & \frac{pD(p)S_{-i}^{\epsilon_1}}{(S_{-i}^{\epsilon_1}+s_i^{\epsilon_2})^2}+p\epsilon_2 >\frac{pD(p)S_{-i}^{\epsilon_1}}{(S_{-i}^{\epsilon_1}+s_i^{\epsilon_1})^2}+p\epsilon_1 \\
        \geq & \partial_{+}f_i(x_i^{\epsilon_1})\geq  \partial_{-}f_i(x_i^{\epsilon_2}).
      \end{align*}
      Here the first inequality is given by the fact that function $q(x)=\frac{x}{(x+k)^2}$ is monotone decreasing in $x\geq k$.

  \item If no $i$ satisfies the conditions in above two steps, i.e., $S_{-i}^{\epsilon_2}< S_{-i}^{\epsilon_1}$ and $S_{-i}^{\epsilon_2}< s_{i}^{\epsilon_2}$ for all $i \in\mathcal{K}$.
      Note that $S_{-i}^{\epsilon_2}< s_{i}^{\epsilon_2}$ implies there exists at most one $i$ in $\mathcal{K}$ in this case. Since $\mathcal{K}\neq\emptyset$, there is exactly one $i'\in \mathcal{K}$. By the definition of $\mathcal{K}$, we have $s_j^{\epsilon_2}>s_j^{\epsilon_1}, \, \forall j\neq i$.
      However, we have $S_{-i}^{\epsilon_2}> S_{-i}^{\epsilon_1}$, which gives a contradiction.
\end{enumerate}
In conclusion, we have $S^\epsilon_2(p)\geq S^{\epsilon_1}(p)$, and in particular, $S^\epsilon(p)\geq S^0(p)$.

\subsection{Proof of Theorem \ref{thm:subsidy}}
Suppose $\epsilon>0$ and $S^\epsilon(p)> S^0(p)$. There must exists some $i$ such that $s_i^{\epsilon}> s_i^0$, which means  $x_i^\epsilon < x_i^0$ for such $i$. 
Thus, according to Lemma \ref{lemma:gradient} and (\ref{eq:epsilon_lemma}):
\begin{align*}
\frac{\mathbf{d}Q_i(s_i)}{\mathbf{d}s_i}\bigg|_{s_i^0}
\leq & \partial_{-} f_i(x_i^0) \\
\leq & \partial_{+} f_i(x_i^\epsilon) \\
\leq & \frac{\mathbf{d}Q_i(s_i;p)}{\mathbf{d}s_i}\bigg|_{s_i^\epsilon}+p\epsilon, \nonumber
\end{align*}
Hence, we always have the following inequality:
\begin{align}\label{eq:epsilon_inequality}
 \frac{pD(p)(S^0(p)-s_i^0)}{(S^0(p))^2}&\leq \frac{pD(p)(S^\epsilon(p)-s_i^\epsilon)}{(S^\epsilon(p))^2}+p\epsilon \nonumber\\
 &\leq \frac{pD(p)(S^\epsilon(p)-s_i^0)}{(S^\epsilon(p))^2}+p\epsilon
\end{align}
If $\forall i,\, \epsilon\leq \frac{D(p)}{S^0(p)}(1-\frac{s_i^0}{S^0(p)})$, i.e., the ``small subsidies'' regime in Theorem \ref{thm:subsidy}, the above inequality gives the following result:
\begin{equation}\label{eq:epsilon_result}
 S^\epsilon(p)\leq F(\epsilon)= \frac{D(p)+\sqrt{M^2+4D(p)s_i^0\epsilon}}
 {2[\frac{D(p)}{S^0(p)}(1-\frac{s_i^0}{S^0(p)})-\epsilon]}
\end{equation}
where $M=D(p)(1-\frac{2s_i^0}{S^0(p)})$.
In fact,  one can verify that
\begin{equation}\label{eq:epsilon0_result}
  S^0(p) = F(0)
\end{equation}
Note that, by (\ref{eq:epsilon_result}) and (\ref{eq:epsilon0_result}) we have $S^{\epsilon}(p)\leq S(p)+\Theta(\sqrt \epsilon)$.

For  $\epsilon$ such that $\epsilon> \frac{D(p)}{S^0(p)}(1-\frac{s_i^0}{S^0(p)})=\frac{D(p)(S^0(p)-s_i^0)}{(S^0(p))^2}$,
 the following inequality  always hold:
\begin{align}
  & \frac{pD(p)(S^\epsilon(p)-s_i^\epsilon)}{(S^\epsilon(p))^2}+p\epsilon \nonumber\\
 > & \frac{pD(p)(S^0(p)-s_i^0)}{(S^0(p))^2} \nonumber\\
 \geq & c \nonumber
\end{align}
Hence, we always have $x_i^\epsilon(p)+s_i^\epsilon(p)=1$ (the ``medium subsidies'' regime  in Theorem \ref{thm:subsidy}).

Finally, if $p\epsilon +\frac{pD(p)S_{-i}^{\epsilon}}{(S_{-i}^{\epsilon}+1)^2}\geq \partial_{+}f_i(0)$ (the ``large subsidies'' regime in Theorem \ref{thm:subsidy}), we always have $s_i^\epsilon =1$ since it is always more beneficial to share. Thus, $S^{\epsilon}(p)=N_O$.

\end{document}